\documentclass[12pt, hyper]{article}
\pdfoutput=1
\usepackage{jheppub1,amsmath,amssymb,amsfonts,amsxtra, mathrsfs, makeidx,graphics,graphicx,amsthm,epsfig, youngtab}

\newcommand{\be}{\begin{equation}}
\newcommand{\ee}{\end{equation}}
\newcommand{\beq}{\begin{equation}}
\newcommand{\eeq}{\end{equation}}
\newcommand{\ba}{\begin{array}}
\newcommand{\ea}{\end{array}}
\newcommand{\bi}{\begin{itemize}}
\newcommand{\ei}{\end{itemize}}
\newcommand{\bea}{\begin{eqnarray}}
\newcommand{\eea}{\end{eqnarray}}
\newcommand{\ben}{\begin{enumerate}}
\newcommand{\een}{\end{enumerate}}
\newcommand{\bean}{\begin{eqnarray*}}
\newcommand{\eean}{\end{eqnarray*}}
\newcommand{\eref}[1]{(\ref{#1})}
\newcommand{\sref}[1]{\S\ref{#1}}
\newcommand{\tref}[1]{Table~\ref{#1}}
\newcommand{\fref}[1]{Figure~\ref{#1}}
\newcommand{\nn}{\nonumber}

\newcommand{\tr}{\mathop{\rm Tr}}
\newcommand{\PE}{\mathop{\rm PE}}
\newcommand{\PL}{\mathop{\rm PL}}
\newcommand{\tQ}{\widetilde{Q}}
\newcommand{\tB}{\widetilde{B}}

\newcommand{\BC}{\mathbb{C}}

\newcommand{\BT}{\mathbb{T}}
\newcommand{\BZ}{\mathbb{Z}}

\newcommand{\BU}{\mathbf{1}}

\newcommand{\comment}[1]{}
\newcommand{\CF}{{\cal F}}

\newcommand{\CM}{{\cal M}}

\newcommand{\CN}{{\cal N}}

\newcommand{\diag}{\mathrm{diag}}

\newcommand{\ie}{{\it i.e.}}
\newcommand{\eg}{{\it e.g.}}

\newcommand{\ud}{\mathrm{d}}
\newcommand{\tti}{\widetilde{t}}

\newcommand{\tV}{\widetilde{V}}

\newtheorem{theorem}{\bf Theorem}

\title{The Hilbert series of U/SU SQCD and Toeplitz Determinants}

\author[a]{Yang Chen}
\author[b]{and Noppadol Mekareeya} 

\affiliation[a]{Department of Mathematics, Imperial College London, \\
180 Queen's Gates, London SW7 2BZ, UK}
\affiliation[b]{Theoretical Physics Group, The Blackett Laboratory \\
Imperial College London, Prince Consort Road\\ 
London,  SW7 2AZ,  UK}
\emailAdd{ ychen@imperial.ac.uk}
\emailAdd{n.mekareeya07@imperial.ac.uk}
\abstract{We present a new technique for computing Hilbert series of $\CN=1$ supersymmetric QCD in four dimensions with unitary and special unitary gauge groups. We show that the Hilbert series of this theory can be written in terms of determinants of Toeplitz matrices. Applying related theorems from random matrix theory, we compute a number of exact Hilbert series as well as asymptotic formulae for large numbers of colours and flavours -- many of which have not been derived before.}
\begin{document}
\maketitle

\section{Introduction and summary}
Given a supersymmetric gauge theory, the space of solutions of the vacuum equations, known as the {\it moduli space of vacua} or simply {\it the moduli space}, is one of the first properties one can study. The phase structure and the possible excitations at a given vacuum configuration can be investigated. In supersymmetric gauge theories, holomorphic gauge invariant operators play a central role in giving us information about the structure of the moduli space. It is therefore worthwhile to study and count these holomorphic functions. This can be done in a similar fashion to statistical mechanics, \ie~by constructing a partition function. Mathematically, this partition function is known as the {\it Hilbert series} (see \eg~\cite{Forcella:2009bv, Hanany:2007zz} for reviews of applications of the Hilbert series to gauge theories and string theory). The Hilbert series not only contains information about the spectrum of the operators in the theory, it also carries geometrical properties of the moduli space (see \eg~\cite{Benvenuti:2006qr, Feng:2007ur, Gray:2008yu}). It is also an indicator of whether the moduli space is Calabi-Yau (see \eg~ \cite{Forcella:2008bb,Gray:2008yu,Hanany:2008kn}). Moreover, Hilbert series can be used as a primary tool to test various dualities in gauge theories and in string theory (see \eg~ \cite{Romelsberger:2005eg, Forcella:2008ng, Davey:2009sr, Hanany:2010qu}).

The Hilbert series was first applied to study the moduli space of $\CN=1$ Supersymmetric Quantum Chromodynamics (SQCD) in four dimensions by Pouliot in 1998 \cite{Pouliot:1998yv}. In that paper, a method for computing the Hilbert series using the Molien--Weyl formula was briefly discussed and the Hilbert series of $SU(2)$ SQCD up to 3 flavours were explicitly stated.   In 2005, R\"omelsberger \cite{Romelsberger:2005eg} computed the Hilbert series for $SU(2)$ with 3 flavours and used it to verify the Seiberg duality.  Later, in \cite{Gray:2008yu, Hanany:2008kn}, the Hilbert series were explicitly computed for SQCD with classical gauge groups for various numbers of colours and flavours and from which several geometrical aspects of the moduli space were extracted.  In addition to SQCD, there have been several applications of the Hilbert series to study a wide range of supersymmetric gauge theories and string theory, \eg~  quiver gauge theories on D3-branes \cite{Forcella:2007wk, Butti:2007jv}, $\CN=4$ gauge theory in four dimensions \cite{Dolan:2007rq}, SQCD with one adjoint chiral superfield \cite{Hanany:2008sb}, M2-brane theories \cite{Hanany:2008qc, Hanany:2008fj, Davey:2009sr, Davey:2009qx, Davey:2011mz}, perturbative string spectrum \cite{Hanany:2010da}, moduli spaces of instantons \cite{Nakajima:2003pg, Benvenuti:2010pq}, and a number of $\CN = 2$ gauge theories in four dimensions \cite{Benvenuti:2010pq, Hanany:2006uc, Hanany:2010qu}.

In this paper, we focus on the Hilbert series of SQCD with the gauge groups $U(N_c)$ and $SU(N_c)$ and $N_f$ flavours of fundamental matters.  Similarly to \cite{Gray:2008yu}, the Molien--Weyl formula is used as a starting point of the computations.  However, instead of evaluating the Hilbert series case by case for each $(N_f, N_c)$ as in \cite{Gray:2008yu}, we take another approach of computations.  First, we show that the Molien--Weyl formula for $U(N_c)$ and $SU(N_c)$ SQCD can be rewritten in terms of determinants of Toeplitz matrices.  Then, we evaluate these determinants both {\it exactly} and {\it asymptotically} for a large class of values of $N_f$ and $N_c$ using several theorems from random matrix theory.\footnote{We mention \emph{en passant} that there have been several recent works on applications of matrix models to supersymmetric gauge theories and string theory -- see, for example, \cite{Marino:2004eq, Kapustin:2009kz, Marino:2009jd, Martelli:2011qj, Cheon:2011vi, Jafferis:2011zi, Herzog:2010hf}.}  In particular, such theorems are stated in \sref{BOSzego} for $U(N_c)$ SQCD and in \sref{sec:BWFH} for $SU(N_c)$ SQCD (see \eg~\cite{BottcherWidom} for a mathematical review).  

Since a different method from \cite{Gray:2008yu} is used to compute Hilbert series in this paper, the way our results presented here is generally distinct from that in \cite{Gray:2008yu}. In this paper, the Hilbert series are computed for a class of values of $N_f$ and $N_c$ rather than a specific value of $(N_f, N_c)$ --  for example, the exact Hilbert series for $SU(N_c)$ SQCD with $N_f = N_c$ and $N_f = N_c+1$ are presented respectively in \sref{sec:nfeqsunc} and \sref{sec:nfsuncp1} and such results, of course, apply for any $N_c >1$.  As a consequence, many general results conjectured in \cite{Gray:2008yu} can be proven (or at least can be checked in a non-trivial way) using Toeplitz determinants.  Such a technique also allows us to compute explicity asymptotic formulae for large $N_f$ and $N_c$.  A number of exact and asymptotic results presented in this paper are too difficult or impossible to be derived using the method in \cite{Gray:2008yu}, partly due to a large number of contour integrals in the Molien--Weyl formula and partly due to the complications of the results themselves.

Let us summarise the main results and key features of this paper below.
\subsection*{Outline and key Points:}
\bi
\item In \sref{zerowindingtp}, we give the definitions of a Toeplitz matrix and a Toeplitz determinant. In \sref{BOSzego}, we state important facts, namely the CGBO formula and the strong Szeg\H{o} limit theorem, which are applied to compute Toeplitz determinants for $U(N_c)$ SQCD with $N_f$ flavours.

\item In \sref{sec:unctpdet}, we show that the Hilbert series for $U(N_c)$ SQCD with $N_f$ flavours can be written in terms of a Toeplitz determinant of a symbol with a zero winding number.

\item In \sref{sec:exactunchilb}, we use the CGBO formula to compute exact Hilbert series for $U(N_c)$ SQCD with $N_f \leq N_c$, $N_f = N_c+1$, $N_f =N_c+2$ and $N_f =N_c +3$.  We also conjectured the general expression \eref{mainresult} of the Hilbert series  for any $N_c$ and any $N_f$ written in terms of a sum over representations of $SU(N_f) \times SU(N_f)$.   The information about the moduli space and the chiral ring of $U(N_c)$ SQCD can be extracted from these Hilbert series -- this is summarised in \sref{sec:modulispaceunc}. 

\item In \sref{sec:asympunc}, we study asymptotics for $N_f, N_c >>1$ with a fixed finite difference $N_f- N_c > 0$.  The asymptotic formula for this limit is given by \eref{CGBOdelta}. In \sref{sec:asymprncunc}, we study asymptotics for $N_f, N_c >>1$ with a fixed ratio $N_f/N_c \geq 1$.  The asymptotic formula for this limit is given by \eref{rncunc}.

\item In \sref{sec:sunctpdet}, we show that the Hilbert series for $SU(N_c)$ SQCD with $N_f$ flavours can be written in terms of a sum of 3 parts: (i) the Hilbert series of $U(N_c)$ SQCD with $N_f$ flavours, (ii) Toeplitz determinants of symbols with positive winding numbers, and (iii) Toeplitz determinants of symbols with negative winding numbers.  Observe that we can use previous results from $U(N_c)$ SQCD in part (i) of the $SU(N_c)$ SQCD Hilbert series computations. 

\item In \sref{sec:BWFH}, we state important facts, namely the B\"ottcher-Widom theorem and the Fisher-Hartwig theorem, which are applied to compute Toeplitz determinants for $SU(N_c)$ SQCD with $N_f$ flavours.

\item In \S \S \ref{sec:nfleqsuncm1}, \ref{sec:nfeqsunc} and \ref{sec:nfsuncp1}, we use the B\"ottcher--Widom formula to compute exact Hilbert series for $SU(N_c)$ SQCD with $N_f \leq N_c-1$, $N_f = N_c$, $N_f = N_c+1$.  The first two subsections contain the proofs of (3.6), (5.3), (3.15) and (5.2) of \cite{Gray:2008yu}.  These exact results also provide a non-trivial test for the general formula \eref{suncexactref}.

\item In \sref{sec:asympsu}, we apply the Fisher--Hartwig Theorem to examine asymptotics of the Hilbert series of $SU(N_c)$ with $N_f$ flavours when $N_f, N_c >>1$.   In \sref{sec:asympfixeddiffsu}, we study asymptotics for $N_f, N_c >>1$ with a fixed finite difference $N_f- N_c \geq 0$.  The asymptotic formula for this limit is given by \eref{asympdeltasu}. In \sref{sec:asympfixedratio}, we study asymptotics for $N_f, N_c >>1$ with a fixed ratio $N_f/N_c \geq 1$.  The asymptotic formula for this limit is given by \eref{asymprncsunc}.
\ei

\paragraph{Note added.}  In the second version of this paper, the authors become aware of the use of Toeplitz determinants and related theorems in the literature on decaying D-branes; see \eg~ \cite{Balasubramanian:2004fz, Jokela:2005ha, Jokela:2008zh, Jokela:2009fd, Jokela:2010cc}.  

\subsection*{Notation for representations and characters.}
We denote an irreducible representation of a group $SU(r+1)$ by a Dynkin label $[a_1,a_2, \ldots,a_r]$. For example, we denote by $[1,0, \ldots, 0]$ the fundamental representation and by $[0, \ldots,0, 1]$ the anti-fundamental representation. We also use the subscripts $k;L$ and $k;R$ to indicate respectively the $k$-th postitions from the left and the right, \eg~ $1_{k;L}$ in $[0, \ldots, 0, 1_{k;L}, 0, \ldots, 0]$ denotes the 1 in the $k$-th position from the left.  

For representations of the product group $G_1\times G_2$, we use the notation $[\ldots ; \ldots]$ where the tuple to the left of the $;$ is the representation of $G_1$, and the tuple to the right of the $;$ is the representation of $G_2$.  

Since a representation is determined by its character, we denote by $[a_1, a_2, \ldots, a_r]_{x}$ a character of the representation $[a_1, a_2, \ldots, a_r]$ written in terms of the variable $x_1, \ldots, x_r$.  Here the subscript $x$ denotes collectively the set of variables $x_1, \ldots, x_r$.  For example, the characters of the fundamental representation and the anti-fundamental representations of $SU(N_f)$ are given by \eref{fundchar} and \eref{antifundchar}.

\section{$U(N_c)$ SQCD with $N_f$ flavours} \label{sec:uncsqcd}
Consider $\CN=1$ supersymmetric $U(N_c)$ gauge theory in four dimensions with $N_f$ quark $Q^i_a$ and $N_f$ anti-quarks $\tQ^a_i$, where $a = 1, \ldots, N_c$ and $i = 1, \ldots, N_f$.  Let us take the superpotential of this theory to be zero: $W=0$.  The information about the gauge and global symmetries as well as how the matters transform under such symmetries is collected in \tref{thisistableone} (see, \eg, \cite{Shifman:2007kd}).

\begin{table}[htdp]
\begin{center}
{\small
\begin{tabular}{|c||c|cccccc|}
\hline
& Gauge symmetry & & & Global symmetry & & & \\
& $U(N_c) = SU(N_c) \times U(1)$ & $SU(N_f)_1$ & $SU(N_f)_2$ & $U(1)_B$ & $U(1)_R$ & $U(1)_Q$ & $U(1)_{\widetilde{Q} }$\\
\hline \hline
$Q^i_a$ & $[0,\ldots,0,1;-1]$ & $[1,0,\ldots,0]$ & $[0,\ldots,0]$ & 1 & $\frac{N_f-N_c}{N_f}$ & 1 & 0 \\
$\widetilde{Q}^a_i$ & $[1,0,\ldots,0;+1]$ & $[0,\ldots,0]$ & $[0,\ldots,0,1]$ & $-1$ & $\frac{N_f-N_c}{N_f}$ & 0 & $-1$ \\
\hline
\end{tabular}}
\end{center}
\centering \includegraphics[height=1.8cm]{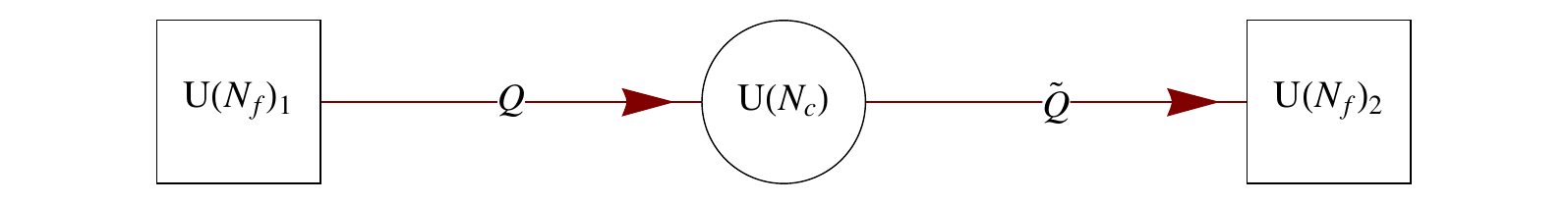}
\caption{The gauge and global symmetries of SQCD and the quantum numbers of the chiral supermultiplets. The quarks are $Q^i_a$ while the antiquarks are $\widetilde{Q}^a_i$. We also draw it as a quiver theory. The circular node represents the $U(N_c)$ gauge symmetry while the two square nodes represent global $U(N_f)_1$ and $U(N_f)_2$ symmetries. Each square node gives rise to a baryonic $U(1)$ global symmetry, one of which is redundant. We thus have $U(1)_{Q, \tilde{Q}}$ that combine into the non-anomalous $U(1)_B$ (sum) and anomalous $U(1)_A$ (difference).}
\label{thisistableone}
\end{table}

$\CN=1$ SQCD with $U(N_c)$ gauge group posesses several interesting phenomena, \eg~ the Seiberg duality \cite{Shifman:2007kd}, non-Abelian flux tubes (strings) and confinements \cite{Shifman:2007kd, Gorsky:2007ip}.  Another motivation for studying this theory is that, as we shall see in \sref{sec:sunctpdet}, Hilbert series for $U(N_c)$ SQCD with $N_f$ flavours are needed for our computations of Hilbert series of $SU(N_c)$ SQCD with $N_f$ flavours.

\subsection{The moduli space of vacua} \label{sec:modulispaceunc}
In this paper, we focus on the classical moduli space of $U(N_c)$ SQCD with $N_f$ flavours. 
\paragraph{The case of $N_f < N_c$.}  At the generic point of the moduli space, the $U(N_c)$ gauge symmetry is partially broken to $U(N_c - N_f)$.  Thus, there are
\bea
N_c^2 - (N_c - N_f)^2 = 2N_c N_f - N_f^2 \nn
\eea
broken generators.  The total number of degrees of freedom of the system is, of course, unaffected by this spontaneous symmetry breaking and the massive gauge bosons each eat one degree of freedom from the chiral matter via the Higgs effect. Therefore, of the original $2N_c N_f$ chiral multiplets (quarks and antiquarks), only 
\bea
2 N_c N_f - (2N_c N_f - N_f^2) = N_f^2 \nn 
\eea
gauge invariant degrees of freedom are left massless. Hence, the dimension of the moduli space is
\bea
\CM_{N_f < N_c} = N_f^2~. 
\eea
We describe these $N_f^2$ massless degrees of freedom by a collection of gauge invariant objects, called {\bf mesons}:
\bea
M^i_j = Q^i_a \tQ^a_j~,
\eea
where the mesons transform in the bi-fundamental representation 
\bea[1,0,\ldots, 0;0,\ldots,0,1]\nn \eea of $SU(N_f)_1 \times SU(N_f)_2$.
The moduli space is said to be {\bf freely generated} by the mesons.

\paragraph{The case of $N_f \geq N_c$.} At the generic point of the moduli space, the $U(N_c)$ gauge symmetry is completely broken. Thus, there are $N_c^2$ broken generators. Therefore, of the original $2N_c N_f$ chiral multiplets (quarks and antiquarks), only 
\bea
2 N_c N_f - N_c^2  \nn 
\eea
gauge invariant degrees of freedom are left massless. Hence, the dimension of the moduli space is
\bea
\dim~\CM_{N_f \geq N_c} = 2 N_c N_f - N_c^2~. \label{UNcNfgeqNc}
\eea
The moduli space is still generated by the mesons:
\bea
M^i_j = Q^i_a \tQ^a_j~.
\eea
However, for $N_f \geq N_c+1$, there are non-trivial relations between these mesons.
The basic relations (which generate the entire ideal of relations) can be written as
\bea
M^{[i_1}_{j_1} \ldots M^{i_{N_c+1}]}_{j_{N_c+1}} = 0~.  \label{relUNc}
\eea
They transform in the representation 
\bea
[0,\ldots,0,1_{N_c+1;L},0, \ldots,0; 0,\ldots,0,1_{N_c+1;R},0, \ldots,0]  \label{reprelUNc}
\eea
of the $SU(N_f)_1 \times SU(N_f)_2$ flavour symmetry.  Note that the dimension of this representation
${N_f \choose N_c+1}^2$ is the number of relations between the mesons.

\paragraph{The case of $N_f = N_c+1$.} In this case, the representation \eref{reprelUNc} reduces to the trivial representation, \ie~ we have precisely one relation.  Equation \eref{relUNc} becomes
\bea
\det M =0~.
%= \frac{1}{N_f!} \epsilon_{i_1 \ldots i_{N_f}}  \epsilon^{j_1 \ldots j_{N_f}} M^{i_1}_{j_1} M^{i_2}_{j_2} \ldots M^{i_{N_f}}_{j_{N_f}} = 0~.
\eea
From \eref{UNcNfgeqNc}, the dimension of moduli space is
\bea
N_f^2 -1~. 
\eea
Since this is the number of generators (which is $N_f^2$) minus the number of relations (which is $1$), the moduli space for the case $N_f = N_c+1$ is a {\bf complete intersection}.

\subsection{The computations of Hilbert series} \label{sec:compUnc}
The Hilbert series of $U(N_c)$ SQCD with $N_f$ flavours can be computed in two steps as follows.

\paragraph{Step 1.} First we consider the space of symmetric functions of quarks $Q$ and antiquarks $\tQ$.  The Hilbert series of this space can be constructed using the {\bf plethystic exponential}, which is a generator for symmetrisation \cite{Benvenuti:2006qr, Feng:2007ur}.  To remind the reader, we define the plethystic exponential of a multi-variable function $g(t_1,...,t_n)$ that vanishes at the origin, $g(0,...,0) = 0$, to be
\bea
\PE [ g(t_1, \ldots, t_n) ] := \exp \left( \sum_{r=1}^\infty \frac{1}{r} g(t_1^r, \ldots, t_n^r) \right)~.
\eea

Let $t$ be the $U(1)_Q$ global charge fugacity and $\tti$ be the $U(1)_{\tQ}$ global charge fugacity.  Note that $0 \leq t, \tti < 1$.  Let $z_1, \ldots, z_{N_c}$ be the fugacities of the $U(N_c)$ gauge group.  The character of the fundamental representation $[1, 0, \ldots, 0]_{+1}$ of $U(N_c)$ can be written as $\sum_{a=1}^{N_c} z_a$, whereas the character of the antifundamental representation $[0, \ldots, 0,1]_{-1}$ of $U(N_c)$ can be written as $\sum_{a=1}^{N_c} z^{-1}_a$  

Let $x_1, \ldots, x_{N_f}$ be the $SU(N_f)_1$ fugacities and let $y_1, \ldots, y_{N_f}$ be the $SU(N_f)_2$ fugacities. Explicitly, we take the character of the fundamental representation of $SU(N_f)_1$ to be
\bea
[1,0,\ldots,0]_x  = x_1 + \sum_{k=1}^{N_f-2} \frac {x_{k+1}} {x_{k} } + \frac {1} {x_{N_f-1} }~, \label{fundchar}
\eea
and take the character of the anti-fundamental representation of $SU(N_f)_2$ to be
\bea
[0,0,\ldots,1]_y = \frac{1} {y_1}  + \sum_{k=1}^{N_f-2} \frac {y_{k} } {y_{k+1}} + y_{N_f-1}~. \label{antifundchar}
\eea

Then, the Hilbert series of the space of symmetric functions of quarks $Q$ and antiquarks $\tQ$ can be written as
\bea
 \PE \left[ t [1,0,\ldots,0]_x \sum_{a=1}^{N_c} z_a^{-1} + \tti [0,0,\ldots,1]_y \sum_{a=1}^{N_c} z_a\right]~, \label{C2NcNfHS}
\eea
where $t [1,0,\ldots,0]_x \sum_{a=1}^{N_c} z_a^{-1}$ comes from the quarks $Q$ and $\tti [0,0,\ldots,1]_y \sum_{a=1}^{N_c} z_a$ comes from the antiquarks $\tQ$.
Using the definition of the plethystic exponential and the identity $-\log(1-x) = \sum_{k=1}^\infty x^k/k$, we can write down the expression \eref{C2NcNfHS} explicitly as a rational function:
\bea
&& \prod_{a=1}^{N_c}  \frac{1}{\left(1- t z_a^{-1} x_1 \right) \prod_{k=1}^{N_f-2} \left(1- t z_a^{-1} \frac{x_{k+1}}{x_{k}} \right) \left(1- t z_a^{-1} x_{N_f-1} \right)} \nn \\
&& \times \prod_{a=1}^{N_c} \frac{1}{\left(1- \tti z_a \frac{1}{y_1} \right) \prod_{k=1}^{N_f-2} \left(1- \tti z_a \frac{y_k}{y_{k+1}} \right) \left(1- \tti z_a y_{N_f-1} \right)}~.
\eea
One may set $x_1, \ldots, x_{N_c}, y_1, \ldots, y_{N_c}$ to unity (this process is called an {\bf unrefinement}) and obtain
\bea
\prod_{a=1}^{N_c} \frac{1}{(1-\frac{t}{z_a})^{N_f} (1-\tti z_a)^{N_f} }~.
\eea

\paragraph{Step 2.} Since the moduli space is parametrised by gauge invariant quantities, we need to project representations associated with symmetric functions in $Q$ and $\tQ$ discussed in Step 1 onto the trivial subrepresentation, which consists of the quantities invariant under the action of the gauge group.  Using knowledge from representation theory (known as the {\bf Molien-Weyl formula} -- see \eg~\cite{DK, Djokovich}), this can be done by integrating over the whole gauge group.  In this section, we are interested in the $U(N_c)$ gauge group, whose Haar measure is given by (see \eg~\cite{FH,KS})
\bea
\int \ud \mu_{U(N_c)}  &=& \frac{1}{N_c! (2 \pi i)^{N_c}} \oint_{|z_1| =1}  \frac{\ud z_1}{z_1} \cdots \oint_{|z_{N_c}| =1}  \frac{\ud z_{N_c}}{z_{N_c}} | \Delta_{N_c} (z)|^2~, \label{UNHaar}
\eea
where
\bea
| \Delta_{N_c} (z)|^2  = \prod_{1 \leq a <b \leq N_c} |z_a - z_b|^2 ~. \label{vander}
\eea
The Hilbert series for $U(N_c)$ SQCD with $N_f$ flavours is given by
\bea \label{UNcNfref}
&& g_{N_f, U(N_c)} (t, \tti,x,y) = \int \ud \mu_{U(N_c)}   \PE \left[ t [1,0,\ldots,0]_x \sum_{a=1}^{N_c} z_a^{-1} + \tti [0,0,\ldots,1]_y \sum_{a=1}^{N_c} z_a\right] ~, \nn \\
\eea
where here and henceforth we write $x$ and $y$ as collective notation for $x_1, \ldots, x_{N_c}$ and $y_1, \ldots,y_{N_c}$.
Setting $x_1, \ldots, x_{N_c}$ and $y_1, \ldots,y_{N_c}$ to unity, we obtain the unrefined Hilbert series of $U(N_c)$ SQCD with $N_f$ flavours:
\bea
g_{N_f, U(N_c)} (t, \tti)= \int \ud \mu_{U(N_c)}\prod_{a=1}^{N_c} \frac{1}{(1-\frac{t}{z_a})^{N_f} (1-\tti z_a)^{N_f} }~. \label{UNcNf}
\eea

\subsubsection{Toeplitz matrices and Toeplitz determinants} \label{zerowindingtp}
In this paper, we are concerned with computing the integrals \eref{UNcNfref} and \eref{UNcNf} by means of Toeplitz matrices and their determinants.   In this section, let us briefly summarise important facts which will be useful in subsequent computations.  We follow \cite{BottcherWidom} closely and neglect subtleties regarding convergences.  We also refer the reader to \cite{BottSil} for a comprehensive review on Toeplitz operators.

Let $\BT$ be a complex unit circle  and let $\phi: \BT \rightarrow \BC$ be a continuous function such that $\phi(z) \neq 0$ for all $z \in \BT$ and $\phi$ has a zero winding number around zero, \ie
\bea
0 = \frac{1}{2 \pi i} \oint_{|z|=1} \frac{\ud z}{z} \frac{\phi'(z)}{\phi(z)}~. 
\eea
The Fourier coefficients $\phi_k$ (with $k\in \BZ$) of $f$ is defined by
\bea
\phi_k := \frac{1}{2 \pi} \int_0^{2 \pi} \ud \theta f(e^{i\theta}) e^{-ik \theta} = \frac{1}{2 \pi i} \oint_{|z|=1} \frac{\ud z}{z}~z^{-k} f(z)~.
\eea
The $n \times n$ {\bf Toeplitz matrix} is defined by
\bea
T_n (\phi) := \left( \phi_{j-k} \right)_{j,k=1}^n~. \label{def:Toeplitzmat}
\eea
The function $\phi$ is sometimes referred to as the {\bf symbol} of the Toeplitz matrix $T_n$.  We are also interested in computing the determinant of the Toeplitz matrix,
\bea
D_n (\phi) := \det T_n (\phi)~. 
\eea
This is called the {\bf Toeplitz determinant} of the symbol $\phi$.  We also define the {\bf infinite Toeplitz matrix} $T(\phi)$ and  the {\bf infinite Hankel matrix} $H(\phi)$ as
\bea
T (\phi) := \left( \phi_{j-k} \right)_{j,k=1}^\infty~, \qquad H(\phi) :=  \left( \phi_{j+k-1} \right)_{j,k=1}^\infty~.
\eea
Subsequently, we find that the following definition is useful
\bea
H(\widetilde{\phi}) :=  \left( \phi_{-j-k+1} \right)_{j,k=1}^\infty~.
\eea

\subsubsection{The Geronimo-Case-Borodin-Okounkov (GCBO) formula and the strong Szeg\H{o} limit theorem} \label{BOSzego}
There are theorems which are very useful for computing Toeplitz determinants (see, \eg, \cite{BasorWidom, BottcherWidom}). Before stating these theorems, let us give more definitions which will be extensively referred to later.

Under general conditions $\phi$ can be written as 
\bea 
\phi = \phi_+ \phi_-~,
\eea 
where $\phi_+$ (resp. $\phi_-$) is a nonzero analytic function in the interior (resp. exterior) of $\BT$.  
This is factorisation is known as the {\bf Wiener-Hopf factorisation}.
Let us define the functions
\bea
U := \frac{\phi_-}{\phi_+}~, \qquad V := \frac{\phi_+}{\phi_-}~.
\eea
Define the Fourier coefficients of $U$ and $V$ as 
\bea
U_{n} := \frac{1}{2 \pi i} \oint_{|z|=1}  \ud z \left(\frac{\phi_-}{\phi_+} \right)  z^{-n-1}, \qquad V_{n} := \frac{1}{2 \pi i} \oint_{|z|=1}  \ud z \left(\frac{\phi_+}{\phi_-} \right)  z^{-n-1} ~.  \label{UVdef}
\eea
We also define
\bea
G(\phi) := \exp (\log \phi)_0 = \exp \left( \frac{1}{2 \pi i} \oint_{|z|=1}  \ud z ~z^{-1} \log \phi \right)~. \label{Gphi}
\eea
Define the projection matrices $P_k$ and $Q_k$ to be infinite matrices such that 
\bea \label{defPQ}
P_k = \diag(\underbrace{1,1, \ldots, 1}_{k~\text{ones}},0,0,\ldots)~, \quad Q_k = \diag(\underbrace{0,0, \ldots, 0}_{k~\text{zeros}},1,1,\ldots)~. 
\eea

Now let us state the theorems.  For the proofs, we refer the reader to \cite{BottcherWidom}.
\begin{theorem} \label{thm:BO} {\bf (The Geronimo-Case-Borodin-Okounkov (GCBO) formula. \cite{Geronimo:1979iy, ref:BO})} The operator $\BU - H(b) H(\widetilde{c})$ is invertible and the $n\times n$ Toeplitz determinant of the symbol $\phi$ is given by
\bea
D_n (\phi) &=& G(\phi)^n E(\phi) \det(\BU - Q_n H(b) H(\widetilde{c}) Q_n) \nn \\
&=& G(\phi)^n E(\phi) \det(\BU - K_n)~,
\eea
where the $(i,j)$-entry of the infinite matrix $K_n$ is
\bea \label{KNc}
K_{n} (i,j) = \left \{ \begin{array}{ll}  \sum_{k=1}^\infty U_{i+k} V_{-j-k} & \text{if~ $i,j \geq n$}  \\ 
 0 & \text{otherwise}~, \end{array} \right.
\eea
and 
\bea
E(\phi) &=& \exp \left( \sum_{k=1}^\infty k (\log \phi)_k (\log \phi)_{-k} \right)~.   \label{Ephi}
\eea 
\end{theorem}

\begin{theorem} \label{thm:szego} {\bf (The Szeg\H{o} strong limit theorem. \cite{ref:szego})} In the limit of large $n$, we have
\bea
D_n(\phi) \sim G(\phi)^n E(\phi)~,
\eea 
where $G(\phi)$ is defined as in \eref{Gphi} and $E(\phi)$ is defined as in \eref{Ephi}~.
\end{theorem}

In the following subsections, we give explicit examples on how to apply these theorems to compute Hilbert series of $U(N_c)$ SQCD with $N_f$ flavours.

\subsubsection{The Hilbert series and the Toeplitz determinant} \label{sec:unctpdet}
Now let us go back to our problem of computing \eref{UNcNfref} and \eref{UNcNf}.  Let 
\bea
\phi(t, \tti, x, y, z) = \PE\left[ t [1,0,\ldots,0]_x z^{-1} + \tti [0,0,\ldots,1]_y z\right]~.
\eea
Note that $\phi(t, \tti, x, y, z)$ is non-zero for all $z \in \BT$ and it has a zero winding number around $z=0$. The Hilbert series can be written as
\bea
 g_{N_f, U(N_c)} (t, \tti,x,y) &=&  \int \ud \mu_{U(N_c)} (z_1, \ldots, z_{N_c}) \prod_{a=1}^{N_c} \phi(t, \tti, x, y, z_a)~. \label{haarsym}
\eea
It is well-known that the $U(N_c)$ Haar measure can be written in terms of a determinant.  For the sake of completeness and ease of reading, we include a proof in Appendix \ref{app:haardet}. In particular, from \eref{haaruncdet}, we find that 
\bea
\int \ud \mu_{U(N_c)}  = \det \left( \frac{1}{2 \pi i} \oint_{|z| =1} \frac{\ud z}{z}  z^{a-b} \right) _{1 \leq a,b \leq N_c}~,
\eea
and, from \eref{gramwithphi}, we have 
%Regarding $\prod_{a=1}^{N_c} \phi(t, \tti, x, y, z_a)$ as a determinant of an $N_c \times N_c$ diagonal matrix whose diagonal entries are $\phi(t, \tti, x, y, z_a)$, we have
\bea
 g_{N_f, U(N_c)} (t, \tti,x,y)  &=& \det \left( \frac{1}{2 \pi i} \oint_{|z| =1} \frac{\ud z}{z}  z^{a-b} \phi(t, \tti, x, y, z) \right)_{1 \leq a,b \leq N_c} \nn\\
&=& D_{N_c} (\phi)~. \label{HSD}
\eea
This is simply the $N_c \times N_c$ Toeplitz determinant with the symbol $\phi(t, \tti, x, y, z)$.  We therefore take $n$ in Theorem \ref{thm:BO} and Theorem \ref{thm:szego} to be the number of colours $N_c$.

\paragraph{Results for any $N_f$ and any $N_c$.} Before we proceed, let us state some useful results for future references.  The following statements are true for any $N_f$ and any $N_c$,
\bea
G(\phi) &=& \exp \left( \frac{1}{2 \pi i} \oint_{|z|=1}  \frac{\ud z}{z} ~ \log \phi (t, \tti,x,y,z) \right) = 1~, \label{Gphi1}\\
E(\phi) &=& \PE \left[ [1,0, \ldots, 0;0, \ldots, 0,1]_{x;y} t \tti \right]~, 
\eea
where the first equality can be easily checked and the second equality will be proven in the next subsection.

Let us now look at various cases of $N_f$ and $N_c$.

\subsection{The strong Szeg\H{o} limit theorem for the $N_c >> N_f$ limit}
In the case, it follows from the strong Szeg\H{o} limit theorem that
\bea
\log g_{N_c >> N_f} (t, \tti,x,y) \sim \sum^\infty_{k=1} k (\log \phi)_k (\log \phi)_{-k}~,
\eea
where, for $n \in \BZ$, $(\log \phi)_n$ is defined to be the $n$-th Fourier coefficient of $\log \phi$:
\bea
(\log \phi)_n := \frac{1}{2 \pi i} \oint_{|z|=1}  \ud z   z^{-n-1} \log \phi~.
\eea
This can be evaluated exactly as
\bea
(\log \phi)_n  = \left \{ \begin{array}{ll}  \frac{1}{n} [0,\ldots,0,1]_{y^n} \tti^n \quad & \text{if~ $n>0$} \\ 
0 & \text{if~ $n=0$} \\
\frac{1}{|n|} [1,0,\ldots,0]_{x^{|n|}} t^{|n|} \quad & \text{if~ $n<0$}  ~. \end{array} \right.
\eea
where $[0,\ldots,0,1]_{y^n} = \frac{1} {y_1^n}  + \frac{y_2^n}{y_1^n} + \ldots \frac{ y_{N_f-2}^n}{ y_{N_f-1}^n}+ y_{N_f-1}^n$, and similarly for $[1,0,\ldots,0]_{x^{|n|}}$.
Hence, it follows from \eref{Ephi} that
\bea
E(\phi) &=& \exp \left( \sum^\infty_{k=1} \frac{1}{k} [1,0, \ldots, 0;0, \ldots, 0,1]_{x^k;y^k} (t \tti)^k \right) \nn \\
&=& \PE \left[ [1,0, \ldots, 0;0, \ldots, 0,1]_{x;y} t \tti \right]~. \label{szeref}
\eea
It follows immediately from Theorem \ref{thm:szego} and \eref{Gphi1} that
\bea
g_{N_c >> N_f} (t, \tti,x,y) \sim E(\phi) = \PE \left[ [1,0, \ldots, 0;0, \ldots, 0,1]_{x;y} t \tti \right]~.
\eea
This formula can also be written in terms of sums of irreducible representations as
\bea
&& g_{N_c >> N_f} (t, \tti,x,y) \sim \frac{1}{1-(t \tti)^{N_f}} \times \nn \\
&& \sum_{n_1, \ldots, n_{N_f-1} = 0}^\infty [n_1, n_2, \ldots, n_{N_f-1}~;~n_{N_f-1}, \ldots, n_2,n_1] (t \tti)^{\sum_{j=1}^{N_f-1} j n_j}~. 
\eea

\paragraph{The unrefined Hilbert series.} Setting $x_1, \ldots, x_{N_c}$ and $y_1, \ldots y_{N_c}$ to unity in \eref{szeref}, we replace the representations by their dimensions. Note that
\bea
\dim [1,0, \ldots, 0] = \dim [0, \ldots, 0,1] = N_f~.
\eea
Hence, we have the unrefined Hilbert series: 
\bea
g_{N_c >> N_f} (t, \tti,1, 1) \sim \PE[N_f^2~t \tti ] = \frac{1}{(1- t \tti)^{N_f^2}}~. \label{aympNcggNf}
\eea

In the next subsection, we use the GCBO formula to show that {\bf \eref{szeref} and \eref{aympNcggNf} are actually exact for $N_f \leq N_c$}.  In other words, corrections to the asymptotic formula \eref{szeref} and \eref{aympNcggNf} are indeed zero for $N_f \leq N_c$.

\subsection{The GCBO formula and exact results} \label{sec:exactunchilb}
We perform the Wiener-Hopf factorisation of $\phi$ as
\bea
\phi(t, \tti, x,y,z) = \phi_+(t,y,z) \phi_-(\tti,x,z)~,
\eea
where
\bea
\phi_{+}(t,y,z) = \PE \left[ \tti [0,0,\ldots,1]_y z  \right], \qquad \phi_-(\tti,x,z) = \PE \left[ t [1,0,\ldots,0]_x z^{-1} \right]~. \nn \\
\eea
%Observe that $\phi_+$ (resp. $\phi_-$) is a non-zero analytic function in the interior (resp. exterior) of the unit circle, and the symbol $\phi$ is simply the product of $\phi_+$ and $\phi_-$:
%\bea
%\phi(t, \tti, x,y,z) = \phi_+(t,y,z) \phi_-(\tti,x,z)~. 
%\eea
%Now let us define a matrix $K_{N_c}$ such that the $(i,j)$-entry is given by
%\bea \label{KNc}
%K_{N_c} (i,j) = \left \{ \begin{array}{ll}  \sum_{k=1}^\infty U_{i+k} V_{-j-k} & \text{if~ $i,j \geq N_c$}  \\ 
 %0 & \text{otherwise}~, \end{array} \right.
%\eea
%where
%\bea
%U_{n} = \oint_{|z|=1}  \ud z \left(\frac{\phi_-}{\phi_+} \right)  z^{-n-1}, \quad V_{n} = \oint_{|z|=1}  \ud z \left(\frac{\phi_+}{\phi_-} \right)  z^{-%n-1} ~. 
%\eea
Using \eref{UVdef}, we find that $U_n$ and $V_n$ can be evaluated exactly.  For $0 \leq n \leq N_f$,
\bea
U_n &=& (-\tti)^{n}\sum_{m=0}^{N_f -n} [m,0, \ldots,0~;~0, \ldots, 0, 1_{(n+m);R},0, \ldots,0]_{x; y} (-t \tti)^{m}~, \qquad \nn \\
V_{-n} &=& (-t)^{n}\sum_{m=0}^{N_f -n} [0, \ldots, 0, 1_{(n+m);L},0, \ldots,0~;~0, \ldots,0,m]_{x; y} (-t \tti)^{m}~, \label{UVref}
\eea
where the subscripts ${(n+m);L}$ and ${(n+m);R}$ indicate respectively the $(n+m)$-th position from the left and the right.
For $n \geq N_f+1$, we have
\bea
U_{n} = V_{-n}=  0~. \label{Uvanish}
\eea 

\paragraph{Unrefined results.} Setting $x_1, \ldots, x_{N_c}$ and $y_1, \ldots y_{N_c}$ to unity in \eref{UVref}, we replace the representations by their dimensions. 
Note that
\bea
\dim [m,0, \ldots,0] &=& \dim [0, \ldots,0,m]= {N_f +m -1 \choose m}~, \nn \\
\dim [0, \ldots, 0, 1_{(n+m);L},0, \ldots,0] &=& \dim [0, \ldots, 0, 1_{(n+m);R},0, \ldots,0] = {N_f \choose n+m}~. \qquad
\eea
For $0 \leq n \leq N_f$, they can be written in terms of hypergeometric functions:
\bea
U_n &=& (-\tti)^{n}\sum_{m=0}^{N_f -n} {N_f +m -1 \choose m}{N_f \choose n+m} (-t \tti)^{m} \qquad \nn \\
&=&  (-\tti)^{n}  {N_f \choose n} {}_2F_1 ( n-N_f, N_f; n+1; t \tti)~, \label{un}  \\
V_{-n} &=& (-t)^{n}\sum_{m=0}^{N_f -n} {N_f +m -1 \choose m}{N_f \choose n+m} (-t \tti)^{m} \nn \\
&=&  (-t)^{n}  {N_f \choose n} {}_2F_1 ( n-N_f, N_f; n+1; t \tti)~.\label{vn}
%U_{n} &=&  (-t)^{n} (1-t \tti) \sum_{m=0}^{N_f -n-1} {N_f +m \choose N_f -n} {N_f -n-1 \choose m} (-t \tti)^m \nn \\
%&=& (-t)^{n} (1-t \tti) {N_f \choose n} {}_2F_1 ( n+1-N_f, N_f+1, n+1, t \tti)~, \label{un}  \\
%V_{-n} &=&  (-\tti)^{n} (1-t \tti) \sum_{m=0}^{N_f -n-1} {N_f +m \choose N_f -n} {N_f -n-1 \choose m} (-t \tti)^m \nn \\
%&=& (-\tti)^{n} (1-t \tti) {N_f \choose n} {}_2F_1 ( n+1-N_f, N_f+1, n+1, t \tti)~. \label{vn}
\eea 
For $n \geq N_f+1$, we have
\bea
U_{n} = V_{-n}=  0~.
\eea 

\paragraph{The exact Hilbert series.} The GCBO formula states that the exact formula for $g_{N_f, U(N_c)}$, for {\bf any} $N_c$ and $N_f$, is given by
\bea
 g_{N_f, U(N_c)}(t, \tti, x, y) =\det( \BU - K_{N_c} ) \PE \left[ [1,0, \ldots, 0;0, \ldots, 0,1]_{x;y} t \tti \right]~.  \label{BOref}
\eea 
Hence, the Hilbert series for any $N_c$ and $N_f$ is simply the Hilbert series for $N_f \leq N_c$ multiplied by a correction factor $\det( \BU - K_{N_c} )$.  In the following subsections, we compute this correction exactly and asymptotically in various cases.

\subsubsection{The case of $N_f \leq N_c$}
Let us apply the GCBO formula to the case of $N_f \leq N_c$.
From \eref{KNc} and \eref{Uvanish}, we have 
\bea
K_{N_c} (i,j) = 0 \quad \text{for all $i,j$}~,
\eea
It is easy to see that
\bea
\det (\BU - K_{N_c}) =  1~.
\eea
Thus, from \eref{BOref}, the refined Hilbert series is given by
\bea \label{HSnfleqNc}
&& g_{N_f \leq N_c} (t, \tti, x, y) = \PE \left[ [1,0, \ldots, 0;0, \ldots, 0,1]_{x;y} t \tti \right] \nn \\
&& = \frac{1}{1-(t \tti)^{N_f}} \sum_{n_1, \ldots, n_{N_f-1} = 0}^\infty [n_1, n_2, \ldots, n_{N_f-1};n_{N_f-1}, \ldots, n_2,n_1] (t \tti)^{\sum_{j=1}^{N_f-1} j n_j}~.\qquad
\eea
The unrefined Hilbert series is
\bea
g_{N_f \leq N_c} (t, \tti, 1, 1) = \frac{1}{(1- t \tti)^{N_f^2}}~. \label{nflequnc}
\eea
Indeed, for $N_f \leq N_c$, the moduli space is of $N_f^2$ dimensional and is freely generated by the mesons.

\subsubsection{The case of $N_f = N_c+1$} 
Let us focus on the case of $N_f = N_c+1$.  Using \eref{KNc}, \eref{UVref} and \eref{Uvanish}, we find that
\bea
K_{N_c} (i,j) = \left \{ \begin{array}{ll} U_{N_c+1}V_{-N_c-1} \qquad & \text{if~ $i= N_c$ and $j = N_c$}  \\ 
 0 & \text{otherwise}~. \end{array} \right.  \nn
\eea
Thus, we have
\bea
K_{N_c} (i,j) = \left \{ \begin{array}{ll} t^{N_f} \tti^{N_f} & \text{if~ $i=j= N_c$ }  \\ 
 0 & \text{otherwise}~, \end{array} \right.
\eea
and so 
\bea
\det (\BU - K_{N_c}) =  1- (t\tti)^{N_f}~.
\eea
Thus, from \eref{BOref}, the refined Hilbert series is given by
\bea \label{nfncp1charexp}
&& g_{N_f, U(N_f-1)} (t, \tti, x, y) = ( 1- (t\tti)^{N_f}) \PE \left[ [1,0, \ldots, 0;0, \ldots, 0,1]_{x;y} t \tti \right] \nn \\
&& = \sum_{n_1, \ldots, n_{N_f-1} = 0}^\infty [n_1, n_2, \ldots, n_{N_f-1};n_{N_f-1}, \ldots, n_2,n_1] (t \tti)^{\sum_{j=1}^{N_f-1} j n_j}~.\qquad
\eea
Setting $x$'s and $y$'s to unity, we see that the unrefined Hilbert series is
\bea
g_{N_f, U(N_f-1)} (t, \tti, 1, 1) = \frac{1- (t\tti)^{N_f}}{(1-t \tti)^{N_f^2}}~. \label{unrefnfncp1}
\eea
%Now let us use the identity $\log \left( \det M \right) = \det \left( \log M \right)$:
%\bea
%\log \det (\BU - K_{N_c}) &=& \tr \log (\BU - K_{N_c}) \nn \\
%&=&  - \sum_{n=1}^\infty \frac{1}{n} \tr( K^n_{N_c}) \nn\\
%&=&- \sum_{n=1}^\infty \frac{1}{n} (t^{N_f} \tti^{N_f})^n \nn \\
%&=& \log \left( 1- t^{N_f} \tti^{N_f} \right)~.
%\eea

\subsubsection{The case of $N_f = N_c+2$}
Let us focus on the case of $N_f = N_c+2$.  Using Eq. \eref{KNc} and \eref{Uvanish}, we find that
\bea
K_{N_c} (i,j) = \left \{ \begin{array}{ll} U_{N_c+1}V_{-N_c-1} + U_{N_c+2}V_{-N_c-2} & \text{if~ $i= j =N_c$ }  \\ 
U_{N_c+1}V_{-N_c-2}  & \text{if~ $i  =N_c$, $j= N_c+1$}  \\ 
U_{N_c+2}V_{-N_c-1} & \text{if~ $i= N_c+1$, $j= N_c$}  \\ 
U_{N_c+2}V_{-N_c-2} & \text{if~ $i= N_c+1$, $j= N_c+1$}  \\ 
 0 & \text{otherwise}~. \end{array} \right.  \nn
\eea
Using \eref{UVref}, we find that
\bea
K_{N_c}(N_c, N_c) &=& K_{N_c}(N_f -2, N_f-2)  \nn \\
&=& [1,0, \ldots,0]_x [0,\ldots,0,1]_{y} (t \tti)^{N_f+1} \nn \\
&& - \left( [1,0, \ldots,0,1]_x + [1,0, \ldots,0,1]_y +1 \right) (t \tti)^{N_f} \nn \\
&& +[0, \ldots,0,1]_x [1,0,\ldots,0]_{y} (t \tti)^{N_f-1}~,\nn \\
K_{N_c}(N_c, N_c+1) &=& K_{N_c}(N_f, N_f -1) \nn \\
&=& [1,0, \ldots,0]_x \tti^{N_f} t^{N_f+1} -[1,0, \ldots,0]_y \tti^{N_f-1} t^{N_f} \nn \\
K_{N_c}(N_c+1, N_c) &=& K_{N_c}(N_f-1, N_f) \nn \\
&=& [0,0, \ldots,1]_y t^{N_f} \tti^{N_f+1} -[0,0, \ldots,1]_x t^{N_f-1} \tti^{N_f} \nn \\
K_{N_c}(N_c+1, N_c+1) &=& K_{N_c}(N_f-1, N_f-1) \nn \\
&=&(t \tti)^{N_f}~,
\eea
and $K_{N_c}(i,j) = 0$, otherwise.  Therefore, we obtain
\bea
&& \det(\BU - K_{N_c}) = 1- [0, \ldots,0 ,1; 1,0, \ldots,0]_{x;y} (t \tti)^{N_f-1} \nn \\
&&\qquad + \left( [1,0,\ldots,0,1;0,\ldots,0]_{x;y} +[0,\ldots,0;1,0,\ldots,0,1]_{x;y} \right)(t \tti)^{N_f} \nn \\
&& \qquad - [1,0, \ldots,0 ; 0, \ldots,0,1]_{x;y} (t \tti)^{N_f+1} + (t \tti)^{2N_f}~,
\eea
and the exact Hilbert series for $N_f = N_c+2$ is 
\bea
g_{N_f, U(N_f-2)} (t, \tti, x, y)&=& \PE \left[ [1,0, \ldots, 0;0, \ldots, 0,1]_{x;y} t \tti \right] \times \nn \\
&& \Big[ 1- [0, \ldots,0 ,1; 1,0, \ldots,0]_{x;y} (t \tti)^{N_f-1} \nn \\
&& + \left( [1,0,\ldots,0,1;0,\ldots,0]_{x;y} +[0,\ldots,0;1,0,\ldots,0,1]_{x;y} \right)(t \tti)^{N_f} \nn \\
&&  - [1,0, \ldots,0 ; 0, \ldots,0,1]_{x;y} (t \tti)^{N_f+1} + (t \tti)^{2N_f} \Big]~.
\label{Unfncp2}
\eea
This can be rewritten in terms of a sum of representations of $SU(N_f) \times SU(N_f)$ as
\bea
&& g_{N_f, U(N_f-2)} (t, \tti,x,y) = \sum_{n_1, n_2, \ldots, n_{N_f-2} = 0}^\infty (t \tti)^{\sum_{a=1}^{N_f-2} a n_a} \times \nn \\
&& \qquad\qquad\qquad [n_1,n_2, \ldots, n_{N_f-2}, 0; 0 , n_{N_f-2},\ldots, n_2, n_1]_{x;y}~.
\eea

\paragraph{The unrefined Hilbert series.}  
%Recall from \eref{un} and \eref{vn} that
%\bea
%U_{N_c+1} &=& U_{N_f-1} = N_f (-\tti)^{N_f-1} (1-t \tti) ~, \nn \\
%U_{N_c+2} &=& U_{N_f} = (-\tti)^{N_f}~, \nn \\
%V_{-j-1} &=& 
%\left \{ \begin{array}{ll}  
%N_f (-t)^{N_f-1} (1-t \tti) & \text{if $j =N_c+1 =N_f-1$} \\ 
%(-t)^{N_f} & \text{if $j =N_c+2 = N_f $} \\
%0 & \text{if $j \geq N_c+3 =N_f +1$}~.
%\end{array} \right. \nn
%\eea
Setting $x$'s and $y$'s to unity, we have
\bea
K_{N_c} (i,j) = \left \{ \begin{array}{ll} 
(t \tti)^{N_f} +N_f^2 (t \tti)^{N_f-1} (1-t \tti)^2 \quad & \text{if~ $i=j= N_f-2$ }  \\ 
-N_f \tti^{-1} (t\tti)^{N_f} (1-t \tti) \quad & \text{if~ $i= N_f-2, j =N_f-1$ }\\
-N_f t^{-1} (t \tti)^{N_f } (1-t \tti) \quad & \text{if~ $i= N_f-1, j =N_f-2$ }\\
(t \tti)^{N_f } \quad & \text{if~ $i= j =N_f-1$ }\\
 0 & \text{otherwise}~. \end{array} \right. 
\eea
Hence, we arrive at
\bea
\det( \BU -K_{N_c}) = \left(1-\left(t \tti\right)^{N_f}\right)^2-N_f^2 \left( t \tti\right)^{N_f-1} \left(1- t \tti\right)^2~. 
\eea
The unrefined Hilbert series for the $N_f = N_c+2$ is therefore
\bea
g_{N_f, U(N_f-2)}(t, \tti) = \frac{\left(1-\left(t \tti\right)^{N_f}\right)^2- N_f^2 \left( t \tti\right)^{N_f-1} \left(1- t \tti\right)^2 }{(1-t \tti)^{{N_f}^2}}~.
\eea

\paragraph{Examples.} We list a few examples of unrefined Hilbert series for small $N_f$.
\bea \label{nfnfm2ex}
g_{3, U(1)}(t, \tti) &=& \frac{1+4 t \tilde{t}+t^2 \tilde{t}^2}{\left(1-t \tilde{t}\right)^5}~, \nn \\
g_{4, U(2)}(t, \tti) &=& \frac{1+4 t \tilde{t}+10 t^2 \tilde{t}^2+4 t^3 \tilde{t}^3+t^4 \tilde{t}^4}{\left(1-t \tilde{t}\right)^{12}}~, \nn \\
g_{5, U(3)}(t, \tti) &=& \frac{1+4 t \tilde{t}+10 t^2 \tilde{t}^2+20 t^3 \tilde{t}^3+10 t^4 \tilde{t}^4+4 t^5 \tilde{t}^5+t^6 \tilde{t}^6}{\left(1-t \tilde{t}\right)^{21}}~, \nn \\
g_{6, U(4)}(t, \tti) &=& \frac{1+4 t \tilde{t}+10 t^2 \tilde{t}^2+20 t^3 \tilde{t}^3+35 t^4 \tilde{t}^4+20 t^5 \tilde{t}^5+10 t^6 \tilde{t}^6+4 t^7 \tilde{t}^7+t^8 \tilde{t}^8}{\left(1-t \tilde{t}\right)^{32}}~. \nn \\
\eea

\paragraph{The limit $N_f, N_c >>1$ and $N_f = N_c+2$.}  
In this limit, \eref{Unfncp2} becomes
\bea \label{nfncp2charexp}
&&g_{N_f = N_c+2} (t, \tti, x, y) \sim  \Big[  1- [0, \ldots,0 ,1; 1,0, \ldots,0]_{x;y} (t \tti)^{N_f-1} \nn \\
&&\qquad + \left( [1,0,\ldots,0,1;0,\ldots,0]_{x;y} +[0,\ldots,0;1,0,\ldots,0,1]_{x;y} \right)(t \tti)^{N_f} \nn \\
&& \qquad - [1,0, \ldots,0 ; 0, \ldots,0,1]_{x;y} (t \tti)^{N_f+1} \Big] \PE \left[ [1,0, \ldots, 0;0, \ldots, 0,1]_{x;y} t \tti \right]~.
\eea
Unrefining this Hilbert series, we obtain
\bea
g_{N_f, U(N_f-2)}(t, \tti) \sim \frac{1-N_f^2 (1-t \tti)^2 (t \tti)^{N_f-1}}{(1-t \tti)^{N_f^2}}~. \label{nfncp2unref}
\eea

\subsubsection{The unrefined Hilbert series for $N_f = N_c+3$}
Let us focus on the case of $N_f = N_c+3$.  From \eref{Uvanish}, \eref{un} and \eref{vn}, we find that
\bea
 K_{N_c} (N_c,N_c) &=& (t \tti)^{N_f} +(t \tti)^{N_f-2} \left(1-t \tti\right)^2 \left[{N_f \choose 2}- {N_f +1\choose 2} t \tti\right]^2 \nn \\
 && +(t \tti)^{N_f-1}  \left(1-t \tti\right)^2 N_f^2~, \nn \\
 K_{N_c} (N_c,N_c+1) &=& -\tti^{-1} ( t \tti)^{N_f} \left(1-t \tti\right) N_f \nn \\
 && - \tti^{-1} (t\tti)^{N_f-1} \left(1-t \tti\right)^2\left[{N_f \choose 2}- {N_f +1\choose 2} t \tti\right] N_f~, \nn \\
K_{N_c} (N_c,N_c+2) &=& \tti^{-2} (t\tti)^{N_f} \left(1-t \tti\right)\left[{N_f \choose 2}- {N_f +1\choose 2} t \tti\right] ~, \nn\\
K_{N_c} (N_c+1,N_c+1) &=& \left(t \tti\right)^{-1+N_f} \left[t \tti+\left(1-t \tti\right)^2 N_f^2\right]~, \nn \\
K_{N_c} (N_c+1,N_c+2) &=& -\tti^{-1} (t\tti)^{N_f} \left(1-t \tti\right) N_f~, \nn\\
K_{N_c} (N_c+2,N_c+2) &=& \left(t \tti\right)^{N_f}~, \nn
\eea
and for $N_c+2 \geq i >j \geq N_c$, the entry $K_{N_c} (i,j)$ is simply $K_{N_c} (j,i)$ with $t$ and $\tti$ interchanged. Other entries of $K_{N_c}$ vanish.

Thus, it follows that
{\small
\bea
&& \det (\BU - K_{N_c}) = 1-\left(t \tti\right)^{N_f-2} \left[\left(1-t \tti\right)^2 \left[{ N_f \choose 2}-{ N_f+1 \choose 2}  t \tti\right]^2+t \tti \left(2N_f^2 \left(1-t \tti\right)^2 + 3 t \tti \right) \right]  \nn\\
&& \qquad  +\left(t \tti\right)^{2 N_f-2} \Bigg[ (1-{t \tti})^2 \left({N_f \choose 2}-{t \tti} {N_f +1 \choose 2}\right) \left({N_f \choose 2}-{t \tti} {N_f+1 \choose 2}-2 (1-{t \tti}) N_f^2\right) \nn \\
&& \qquad +N_f^4(1-t \tti)^4 + 2N_f^2 (1-t \tti)^2 t \tti + 3 (t \tti)^2  \Bigg] -\left(t \tti\right)^{3 N_f}~. \label{detnfncp3}
\eea}
Thus, the Hilbert series for the $N_f = N_c+3$ case is 
\bea
g_{N_f, U(N_f-3)}(t, \tti) = \frac{\det (\BU - K_{N_f-3}) }{(1-t \tti)^{{N_f}^2}}~, \label{nfncp3unref}
\eea
where $\det (\BU - K_{N_f-3})$ is given by \eref{detnfncp3}.  

\paragraph{Examples.} We list a few examples of unrefined Hilbert series for small $N_f$.
\bea
g_{4, U(1)} &=& \frac{1+9 t \tilde{t}+9 t^2 \tilde{t}^2+t^3 \tilde{t}^3}{\left(1-t \tilde{t}\right)^7}~, \nn \\
g_{5, U(2)} &=& \frac{1+9 t \tilde{t}+45 t^2 \tilde{t}^2+65 t^3 \tilde{t}^3+45 t^4 \tilde{t}^4+9 t^5 \tilde{t}^5+t^6 \tilde{t}^6}{\left(1-t \tilde{t}\right)^{16}}~, \nn \\
g_{6, U(3)} &=& \frac{1+9 t \tilde{t}+45 t^2 \tilde{t}^2+165 t^3 \tilde{t}^3+270 t^4 \tilde{t}^4+270 t^5 \tilde{t}^5+165 t^6 \tilde{t}^6+45 t^7 \tilde{t}^7+9 t^8 \tilde{t}^8+t^9 \tilde{t}^9}{\left(1-t \tilde{t}\right)^{27}}~. \nn \\
\eea

\paragraph{The limit $N_f, N_c >>1$ and $N_f = N_c+3$.}  
To obtain the asymptotic formula, we approximate $\det (\BU - K_{N_c})$ by neglecting terms of order $t^{2N_f}$ and smaller in comparison with those of order $t^{N_f}$.  Expanding the square bracket in the first line of \eref{detnfncp3} and neglecting terms of order ${N_f \choose 2}$ and smaller in comparison with those of order ${N_f \choose 2}^2$, we obtain
\bea
g_{N_f, U(N_f-3)}(t, \tti) \sim \frac{1- {N_f \choose 2}^2 (1-t \tti)^4  (t \tti)^{N_f-2}}{(1-t \tti)^{N_f^2}}~. \label{asympnfunfm3unref}
\eea

\subsection{An exact refined Hilbert series for any $N_f$ and $N_c$}
From \eref{nfncp1charexp} and \eref{nfncp2charexp}, we see that one can express the Hilbert series in terms of a sum of representations of $SU(N_f) \times SU(N_f)$.  We conjecture that the Hilbert series of $U(N_c)$ SQCD with $N_f$ flavours can be written as
\bea\label{mainresult}
&& g_{N_f, U(N_c)} (t, \tti,x,y) = \sum_{n_1, n_2, \ldots, n_{N_c} = 0}^\infty (t \tti)^{\sum_{a=1}^{N_c} a n_a} \times \nn \\
&& \qquad\qquad\qquad [n_1,n_2, \ldots, n_{N_c}, 0, \ldots,0~;~ 0,\ldots, 0 , n_{N_c},\ldots, n_2, n_1]_{x;y}~.
\eea

\paragraph{A consistency check.} We perform a non-trivial check for this formula by setting $x=y=1$ and obtain an unrefined Hilbert series which can be compared with known results.  Recall the Weyl dimension formula for the $SU(n)$ irreducible representation:
 \bea
 \dim [a_1, \ldots, a_{n-1}] = \prod_{1 \leq i <j \leq n} \frac{(a_i+ \ldots a_{j-1})+(j-i)}{j-i}~.
 \eea

For example, let us take $N_f=5$ and $N_c=3$.  Then,
\begin{equation}
\ba{rcl}
\dim~[n_1,n_2,n_3,0 ; 0 , n_3, n_2, n_1]
&=& \Big[(4!~3!~2!~1!)^{-1}\times \\ \nn
&& (n_1 + 1)(n_1 + n_2 + 2)(n_1 + n_2 + n_3 + 3)(n_1 + n_2 + n_3 + 4)\times \\ \nn
&& (n_2 + 1)(n_2 + n_3 + 2)(n_2 + n_3 + 0 + 3)\times \\ \nn
&& (n_3 + 1)(n_3+ 0 + 2)\times \\ \nn
&& (0 + 1) \Big]^2~.  \label{dim53}
\ea
\end{equation}
From \eref{mainresult}, we find that the unrefined Hilbert series for $N_f=5$ and $N_c=3$ is
\bea
g_{5, U(3)}(t, \tti) &=& \frac{1+4 t \tilde{t}+10 t^2 \tilde{t}^2+20 t^3 \tilde{t}^3+10 t^4 \tilde{t}^4+4 t^5 \tilde{t}^5+t^6 \tilde{t}^6}{\left(1-t \tilde{t}\right)^{21}}~.
\eea
Observe that this coincides with \eref{nfnfm2ex}.

\paragraph{The generators and relations.}  The information about the generators of the moduli space and their relations can be extracted from the Hilbert series using the {\bf plethystic logarithm} \cite{Benvenuti:2006qr, Feng:2007ur}.  To remind the reader, we define the plethystic logarithm of a multi-variable function $g(t_1,...,t_n)$ to be
\bea
\PL [ g(t_1, \ldots, t_n) ] := \sum_{k=1}^\infty \frac{\mu(k)}{k} \log g(t^k_1, \ldots, t^k_n) ~.
\eea
where $\mu(k)$ is the M\"obius function. The significance of the series expansion of the plethystic logarithm is stated in \cite{Benvenuti:2006qr, Feng:2007ur}:  the first terms with plus sign give the generators while the first terms with the minus sign give the relations between these generators.  Now let us compute the plethystic logarithms of various Hilbert series we have computed and interpret the results.  

For $N_f \leq N_c$, we use the Hilbert series is given in the first line of \eref{HSnfleqNc} and so we have
\bea
\PL [g_{N_f \leq N_c} (t, \tti, x, y) ]  = [1,0,\ldots, 0;0,\ldots,0,1]_{x;y} t \tti~.
\eea
This indicates that the moduli space is freely generated by the mesons in the representation $[1,0,\ldots, 0;0,\ldots,0,1]$ of $SU(N_f)\times SU(N_f)$. This agrees with the discussion in \sref{sec:modulispaceunc}.

For $N_f = N_c+1$, we use the Hilbert series is given in the first line of \eref{nfncp1charexp}, and so we have
\bea
\PL [g_{N_c+1, U(N_c)} (t, \tti, x, y) ]  = [1,0,\ldots, 0;0,\ldots,0,1]_{x;y} t \tti - (t \tti)^{N_c+1}~.
\eea
This indicates that the moduli space is a complete intersection.  The generators are the mesons in the representation $[1,0,\ldots, 0;0,\ldots,0,1]$ of $SU(N_f)\times SU(N_f)$, and there is one relation in the trivial representation of $SU(N_f)\times SU(N_f)$. This also agrees with the discussion in \sref{sec:modulispaceunc}.

For $N_f \geq N_c+2$, we use the Hilbert series is given in \eref{mainresult}, and so we have
\bea
&& \PL [g_{N_f \geq N_c+2} (t, \tti, x, y) ]  = [1,0,\ldots, 0;0,\ldots,0,1]_{x;y} t \tti  \nn \\
&& \qquad - [0,\ldots,0,1_{N_c+1;L},0, \ldots,0; 0,\ldots,0,1_{N_c+1;R},0, \ldots,0] (t \tti)^{N_c+1} + \ldots~.
\eea
This indicates that the moduli space is not a complete intersection.  The generators are the mesons in the representation $[1,0,\ldots, 0;0,\ldots,0,1]$ of $SU(N_f)\times SU(N_f)$, and there are relations in the representation $[0,\ldots,0,1_{N_c+1;L},0, \ldots,0; 0,\ldots,0,1_{N_c+1;R},0, \ldots,0]$ of $SU(N_f)\times SU(N_f)$. This also agrees with the discussion in \sref{sec:modulispaceunc}.
  
 \subsection{Various asymptotics} \label{sec:variousasympunc}
In this subsection, we examine asymptotics of Hilbert series for $N_f > N_c$ when $N_f$ and $N_c$ is large.  (Note that for $N_f \leq N_c$, one can simply use the exact formula \eref{nflequnc}.)  

\subsubsection{Asymptotics for $N_f, N_c >>1$ with a fixed difference $N_f- N_c > 0$} \label{sec:asympunc}
Let us focus on the limit of large $N_f$ and $N_c$ where the difference $N_f-N_c$ is kept fixed and being positive.
For convenience, we define
\bea
\Delta := N_f - (N_c+1)~.
\eea
We also assume that $\Delta \geq 0$ and that $\Delta$ is of order $1$.

In this limit, elements of $K_{N_c}$ are much smaller than $1$.  Therefore,
\bea \label{approxdet}
\det (\BU - K_{N_c}) = \exp \left( \tr \log(1-K_{N_c}) \right) \sim \exp \left( - \tr K_{N_c} \right) \sim 1- \tr K_{N_c}~.\qquad
\eea
Now let us consider $\tr K_{N_c}$.  {\bf We claim that the leading contribution to $\tr K_{N_c}$ comes from  $U_{N_c+1} V_{-(N_c+1)}$ in the matrix element $K_{N_c} (N_c, N_c)$.}  Below, we show that for $0 \leq m \leq \Delta-1$,
\bea 
U_{N_c+1+m+1} V_{-(N_c+1+m+1)} \sim N_c^{-2} U_{N_c+1+m} V_{-(N_c+1+m)}~ . 
\eea
Note that from \eref{Uvanish}, for $m \geq \Delta+1$, we have $U_{N_c+1+m} V_{-(N_c+1+m)}=0$.

\begin{proof} Suppose that $ 0 \leq m \leq \Delta-1$.
%\item Let $n = N_c +1+ m=N_f - \Delta +m$. We approximate
%\bea
%{}_2F_1 ( n -N_f, N_f; n+1; t \tti) \sim {}_2F_1 (m- \Delta, N_f; N_f; t \tti) = (1-t \tti)^{\Delta-m}~,
%\eea
%which is of order 1.
Using \eref{un} and \eref{vn}. we obtain
\bea \label{kncm}
U_{N_c+1+m+1} V_{-(N_c+1+m+1)} &=& {N_f \choose \Delta -m-1}^2 \left[{}_2F_1 ( m+1-\Delta, N_f; N_c+2+m+1; t \tti)\right]^2 \times \nn \\
&& (t \tti)^{N_c+1+m+1}~. \nn
\eea
Now consider the binomial coefficient
\bea
{N_f \choose \Delta -m-1}  &=& \frac{N_f!}{(\Delta-m-1)! (N_f - \Delta +m+1)!} \nn \\
&=& \frac{\Delta-m}{N_f - \Delta +m+1} {N_f \choose \Delta-m}\nn \\
&=& \frac{\Delta-m}{N_c+2+m} {N_f \choose \Delta-m} ~. \nn
\eea
Hence, we find that
\bea 
\frac{U_{N_c+1+m+1} V_{-(N_c+1+m+1)}}{U_{N_c+1+m} V_{-(N_c+1+m)}} = (t \tti)  \left[ \frac{\Delta-m}{N_c+2+m}\right]^{2} \left[\frac{{}_2F_1 ( m+1-\Delta, N_f; N_c+2+m+1; t \tti)}{{}_2F_1 ( m-\Delta, N_f; N_c+2+m; t \tti))} \right]^2 ~. \nn
\eea
The ratio between the two hypergeometric functions is of order $1$, \ie~ as $N_c, N_f \rightarrow \infty$,
\bea 
\frac{{}_2F_1 ( m+1-\Delta, N_f; N_c+2+m+1; t \tti)}{{}_2F_1 ( m-\Delta, N_f; N_c+2+m; t \tti))} = O(1)~,
\eea
and so 
\bea \label{UVratio}
\frac{U_{N_c+1+m+1} V_{-(N_c+1+m+1)}}{U_{N_c+1+m} V_{-(N_c+1+m)}} \sim (t \tti)  \left[ \frac{\Delta-m}{N_c+2+m}\right]^{2}~.
\eea
Since $\Delta$ is of order 1, for a large $N_c$ we have
\bea
\frac{U_{N_c+1+m+1} V_{-(N_c+1+m+1)}}{U_{N_c+1+m} V_{-(N_c+1+m)}} \sim \frac{1}{N_c^2}~,
\eea
as claimed.  Thus, in $\tr K_{N_c}$, we may neglect other diagonal elements of $K_{N_c}$ in comparison with $K_{N_c} (N_c, N_c)$, and the leading contribution in $K_{N_c} (N_c, N_c)$ is $U_{N_c+1} V_{-(N_c+1)}$.
\end{proof}
%Therefore, we obtain
%\bea
%U_{N_c+1+m} V_{-(N_c+1+m)} \sim N_f^{-2m} U_{N_c+1} V_{-(N_c+1)}~, \nn
%\eea

%Let us justify the claim using \eref{un} and \eref{vn}.  In what follows, we suppose that $1 \leq m \leq \Delta$.
%\bi
%\item Let $n = N_c +1+ m=N_f - \Delta +m$. We approximate
%\bea
%{}_2F_1 ( n -N_f, N_f; n+1; t \tti) \sim {}_2F_1 (m- \Delta, N_f; N_f; t \tti) = (1-t \tti)^{\Delta-m}~,
%\eea
%which is of order 1.
%\item Therefore, for $N_f, N_c >>1$, we obtain
%\bea \label{kncm}
%U_{N_c+1+m} V_{-(N_c+1+m)} \sim {N_f \choose \Delta -m}^2 (1-t \tti)^{2(\Delta-m)} (t \tti)^{N_f - \Delta+m} ~, \qquad \eea
%which is of order $ {N_f \choose \Delta -m}^2 (t \tti)^{N_f}$.
%\item For $N_f >>1$, we obtain
%\bea
%{N_f \choose \Delta -m}  &=& \frac{N_f!}{(\Delta-m)! (N_f - \Delta +m)!} \nn \\
%&=& {N_f \choose \Delta} \times \frac{\Delta!(N_f - \Delta)!}{(\Delta-m)!(N_f -\Delta+m)!} \sim \frac{1}{N_f^m} {N_f \choose \Delta} ~.
%\eea
%\item Therefore, we obtain
%\bea
%U_{N_c+1+m} V_{-(N_c+1+m)} \sim N_f^{-2m} U_{N_c+1} V_{-(N_c+1)}~, \nn
%\eea
%as claimed.  Thus, in $\tr K_{N_c}$, we may neglect other diagonal elements of $K_{N_c}$ in comparison with $K_{N_c} (N_c+1, N_c+1)$, and the leading contribution in $K_{N_c} (N_c+1, N_c+1)$ is $U_{N_c+1} V_{-(N_c+1)}$.
%\ei

It follows from the above discussion that
\bea \label{apptrknc}
\tr K_{N_c} &\sim& K_{N_c} (N_c+1, N_c+1) \nn \\
&\sim& U_{N_c+1} V_{-(N_c+1)} \nn\\
&=& {N_f \choose \Delta }^2 \left[{}_2F_1 ( -\Delta, N_f; N_c+2; t \tti)\right]^2 (t \tti)^{N_c+1} \nn \\
&\sim& {N_f \choose \Delta}^2  (1-t \tti)^{2 \Delta} (t \tti)^{N_c+1}  ~,
\eea
where we have have used the following approximation to establish the fourth equality:
\bea
{}_2F_1 ( -\Delta, N_f; N_f - \Delta+1; t \tti) \sim {}_2F_1 (- \Delta, N_f; N_f; t \tti) = (1-t \tti)^{\Delta}~.
\eea
Thus, it follows from the CGBO formula that
\bea \label{CGBOdelta}
g_{N_f, U(N_c)}(t, \tti) &\sim& \frac{1- {N_f \choose \Delta}^2  (1-t \tti)^{2 \Delta} (t \tti)^{N_c+1}}{(1-t \tti)^{N_f^2}} ~, \qquad
\eea
for $N_f, N_c >>1$ and the difference $\Delta = N_f - N_c -1 \geq 0$ kept fixed.

Note that for $\Delta = 0$, we recover the exact formula \eref{unrefnfncp1} for $N_f = N_c+1$.  Moreover, for $\Delta =1$ and $\Delta=2$, we recover the asymptotic formulae \eref{nfncp2unref} and \eref{asympnfunfm3unref}, both of which are derived from the exact Hilbert series.

\subsubsection{Asymptotics for $N_f, N_c >>1$ with a fixed ratio $N_f/N_c \geq 1$} \label{sec:asymprncunc}
In this subsection, we focus on the limit $N_f, N_c >> 1$ with a finite ratio 
\bea r := \frac{N_f}{N_c} \geq 1~.\eea  

\paragraph{Derivation of the asymptotic formula.} Let us consider \eref{UVratio}. Since $\Delta = (r-1)N_c-1$, we see that for $0 \leq m \leq \Delta-1$, we have
\bea
\frac{1}{r N_c} \leq \frac{\Delta-m}{N_c+2+m} \leq r-1~.
\eea
Hence, we see that  
\bea  \frac{t \tti}{r^2 N_c^2} \lesssim \frac{U_{N_c+1+m+1} V_{-(N_c+1+m+1)}}{U_{N_c+1+m} V_{-(N_c+1+m)}} \lesssim t \tti (r-1)^2 \eea 
We would like the upper bound to be sufficiently small, so that we have control over $U_{N_c+1+m} V_{-(N_c+1+m)}$ for $m \geq 0$.  We are interested in the following limiting cases:
\bi
\item {\bf Case 1:} Consider $r-1 = o(1)$ as $N_c \rightarrow \infty$.  In other words, we take $r$ to be close to $1$ and $t \tti$ can take any value between $0$ and $1$.
\item {\bf Case 2:} Consider $t \tti = o(1)$ as $N_c \rightarrow \infty$. In other words, we take $t \tti$ to be small and $r$ can take any finite positive value greater than $1$.
\ei
In both cases, we see that $U_{N_c+1+m+1} V_{-(N_c+1+m+1)}$ can be neglected in comparison with $U_{N_c+1+m} V_{-(N_c+1+m)}$.  Thus, the approximation 
\bea
\det (\BU - K_{N_c}) \sim 1- \tr K_{N_c} \sim 1- U_{N_c+1} V_{-(N_c+1)}
\eea
is valid in both limiting cases.
We therefore consider
\bea \label{asympUVunc}
U_{N_c+1} V_{-(N_c+1)}  &=&  {N_f \choose N_c+1}^2  \left[ {}_2F_1 (N_c +1 - N_f, N_f; N_c+2; t \tti) \right]^2 (t \tti)^{N_c+1} \nn \\
&=& {rN_c \choose N_c+1}^2  \left[ {}_2F_1 (-(r-1)N_c+1, r N_c; N_c+2; t \tti) \right]^2 (t \tti)^{N_c+1}~. \qquad
\eea
%Subsequently, we compute the asymptotic formulae for the binomial coefficient and the hypergeometric function.

%The asymptotic formula for the binomial coefficient can be easily computed using the Stirling formula $\log N! \sim N \log N - N+\frac{1}{2} \log(2 \pi N)$ for large $N$, we obtain
%\bea
%\log {r N_c \choose N_c+1} &=& \log (rN_c)! - \log (N_c+1)! - \log \left( (r-1)N_c-1\right)!  \nn \\
%&\sim& N_c \left[ r \log r - (r-1) \log (r-1) \right]+ \frac{1}{2} \log \left( \frac{r(r-1)}{2 \pi N_c} \right)~. 
%\eea
%Therefore, we have
%\bea
%{r N_c \choose N_c+1}  \sim  \frac{1}{\sqrt{2 \pi N_c}} \left[ \frac{r^{N_c r +\frac{1}{2} } }{(r-1)^{N_c(r-1) - \frac{1}{2}}} \right]~.\label{stirling1}
%\eea
Unfortunately, we are not aware of a good asymptotic formula for 
\bea {}_2F_1 ( -(r-1)N_c +1 , r N_c; N_c +2; t \tti)  \nn \eea 
in the limit of large $N_c$, finite $r >1$ and $0< t \tilde{t} <1$. 
Therefore, we leave the expression for the hypergeometric function as it is.\footnote{On the other hand, the asymptotic formula for the binomial coefficient can be easily computed using the Stirling formula $\log N! \sim N \log N - N+\frac{1}{2} \log(2 \pi N)$ for large $N$, we obtain
\bea
\log {r N_c \choose N_c+1} &=& \log (rN_c)! - \log (N_c+1)! - \log \left( (r-1)N_c-1\right)!  \nn \\
&\sim& N_c \left[ r \log r - (r-1) \log (r-1) \right]+ \frac{1}{2} \log \left( \frac{r(r-1)}{2 \pi N_c} \right)~. 
\eea
Therefore, we have
\bea
{r N_c \choose N_c+1}  \sim  \frac{1}{\sqrt{2 \pi N_c}} \left[ \frac{r^{N_c r +\frac{1}{2} } }{(r-1)^{N_c(r-1) - \frac{1}{2}}} \right]~.\label{stirling1}
\eea
However, the asymptotic formula \eref{rncunc} is more elegant when written in terms of the binomial coefficient instead of its asymptotic formula.  Hence, we leave the binomial coefficient as it is.}
 
Therefore, we have
\bea
\det (\BU - K_{N_c})  \sim 1-  {rN_c \choose N_c+1}^2 \left[ \CF(N_c, r, t \tti) \right]^2 (t \tti)^{N_c+1}~,
\eea
where $\CF(N_c, r, t \tti)$ is a shorthand notation for the hypergeometric function:
\bea
\CF(N_c, r, t \tti) := {}_2F_1 ( -(r-1)N_c +1 , r N_c; N_c +2; t \tti)~.
\eea

\paragraph{The asymptotic formula.} Thus, using the GCBO formula, we obtain the required asymptotic formula
\bea \label{rncunc}
g_{rN_c, U(N_c)}(t, \tti) \sim  \frac{1-  {rN_c \choose N_c+1}^2 \left[ \CF(N_c, r, t \tti) \right]^2 (t \tti)^{N_c+1}}{(1- t \tti)^{r^2 N_c^2}}~.
\eea
We emphasise that this asymptotic formula is valid for both limiting cases described above.  We provide non-trivial consistency checks for this formula in the next subsection.

\paragraph{Special case $r=1$:} Observe from \eref{rncunc} that as $r \rightarrow 1$, we have 
\bea
g_{N_c, U(N_c)} (t, \tti) \sim \frac{1}{(1- t \tti)^{N_c^2}}~.
\eea
This indeed coincides with the exact result \eref{nflequnc} for $N_f =N_c$.

\subsubsection{Consistency checks}
In this subsection, we provide consistency checks for the exact and the asymptotic formulae we have derived so far.  

\begin{figure}[htbp]
\begin{center}
\includegraphics[height=6cm]{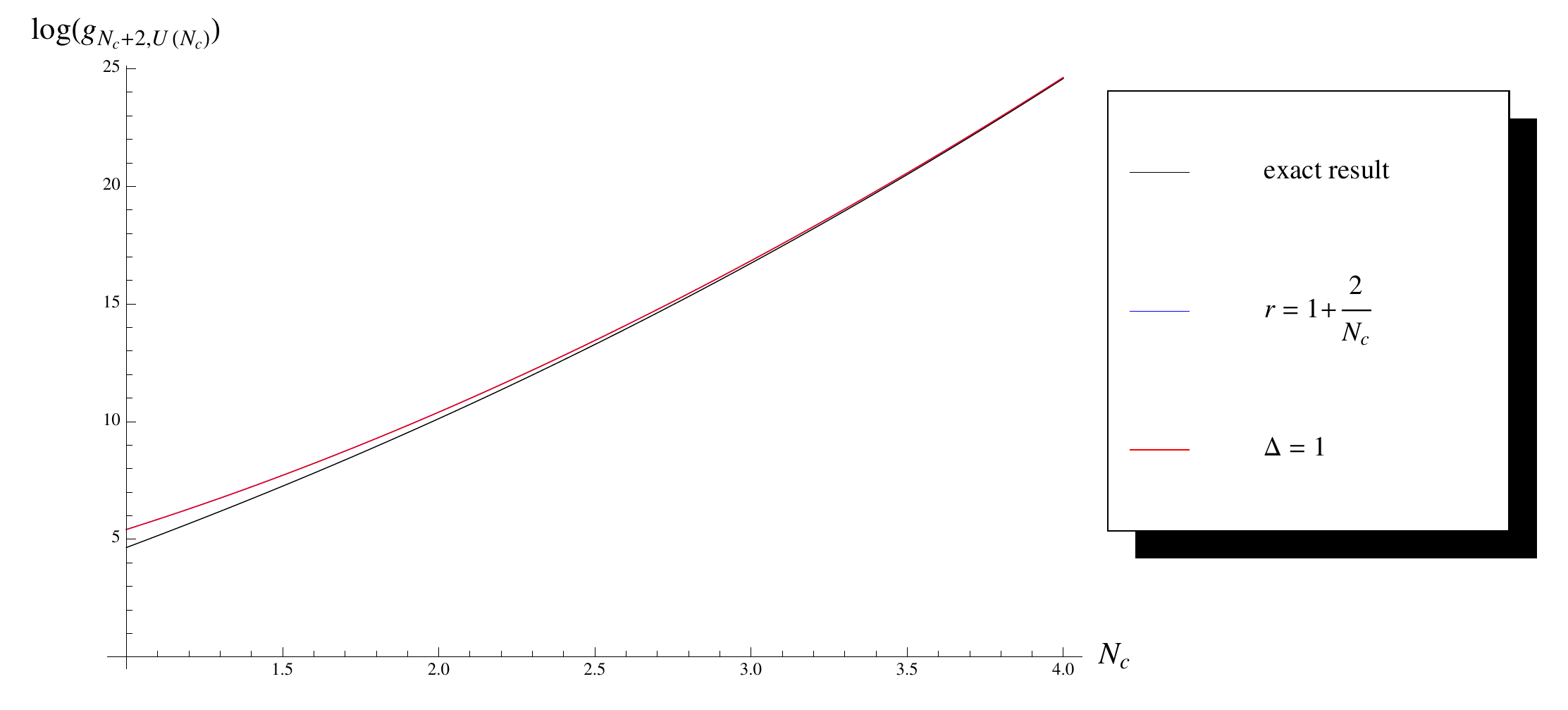}
\caption{The case of $N_f = N_c+2$: The graphs of the exact Hilbert series \eref{nfncp2unref} with its asymptotic formulae \eref{CGBOdelta} and \eref{rncunc}. We take $\Delta =1$, $r=1+ 2/N_c$, $t \tti= 0.5$. Note that the graphs for both asymptotic formulae are on top of each other in this figure.}
\label{fig:ncp2unc}
\end{center}
\end{figure}

In \fref{fig:ncp2unc}, we plot the graphs of $\log g_{N_c+2,U(N_c)}(t, \tti)$ given by \eref{nfncp2unref}, \eref{CGBOdelta} and \eref{rncunc} against $N_c$, with $\Delta =1$, $r=1+ 2/N_c$ and $t \tti = 0.5$.  Note that, for \eref{rncunc}, this is the first limiting case described in \sref{sec:asymprncunc}, since $r-1 \rightarrow 0$ as $N_c \rightarrow \infty$.  It can be see that the asymptotic formulae \eref{CGBOdelta} and \eref{rncunc} approach the exact result \eref{nfncp2unref} when $N_c$ is large.

\begin{figure}[htbp]
\begin{center}
\includegraphics[height=6cm]{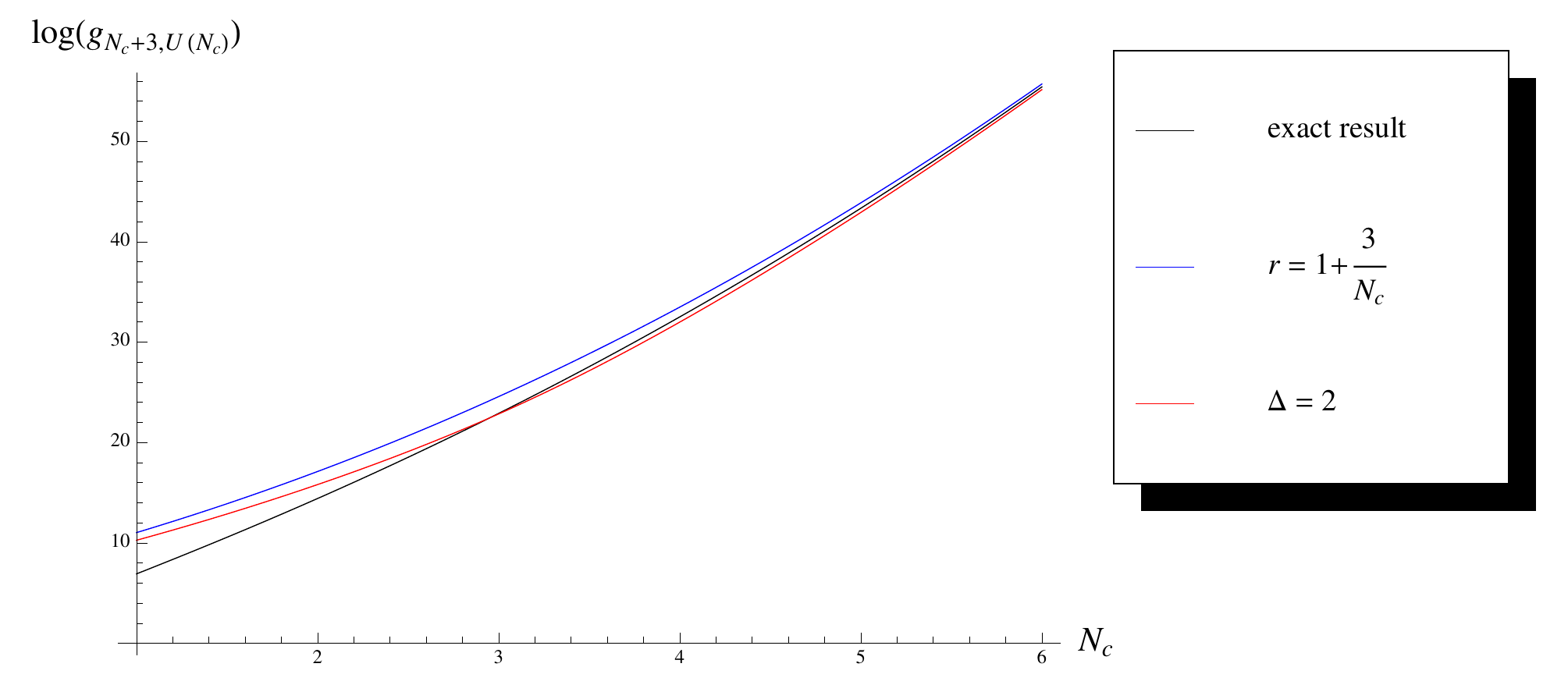}
\caption{The case of $N_f = N_c+3$: The graphs of the exact Hilbert series \eref{nfncp3unref} with its asymptotic formulae \eref{CGBOdelta} and \eref{rncunc}. We take $\Delta =2$, $r=1+ 3/N_c$, $t \tti= 0.5$.}
\label{fig:ncp3unc}
\end{center}
\end{figure}

In \fref{fig:ncp3unc}, we plot the graphs of $\log g_{N_c+3,U(N_c)}(t, \tti)$ given by \eref{nfncp3unref}, \eref{CGBOdelta} and \eref{rncunc} against $N_c$, with $\Delta =2$, $r=1+ 3/N_c$ and $t \tti = 0.5$. Note that, for \eref{rncunc}, this is the first limiting case described in \sref{sec:asymprncunc}, since $r-1 \rightarrow 0$ as $N_c \rightarrow \infty$. It can be see that the asymptotic formulae \eref{CGBOdelta} and \eref{rncunc} approach the exact result \eref{nfncp3unref} when $N_c$ is large.

\begin{figure}[htbp]
\begin{center}
\includegraphics[height=6cm]{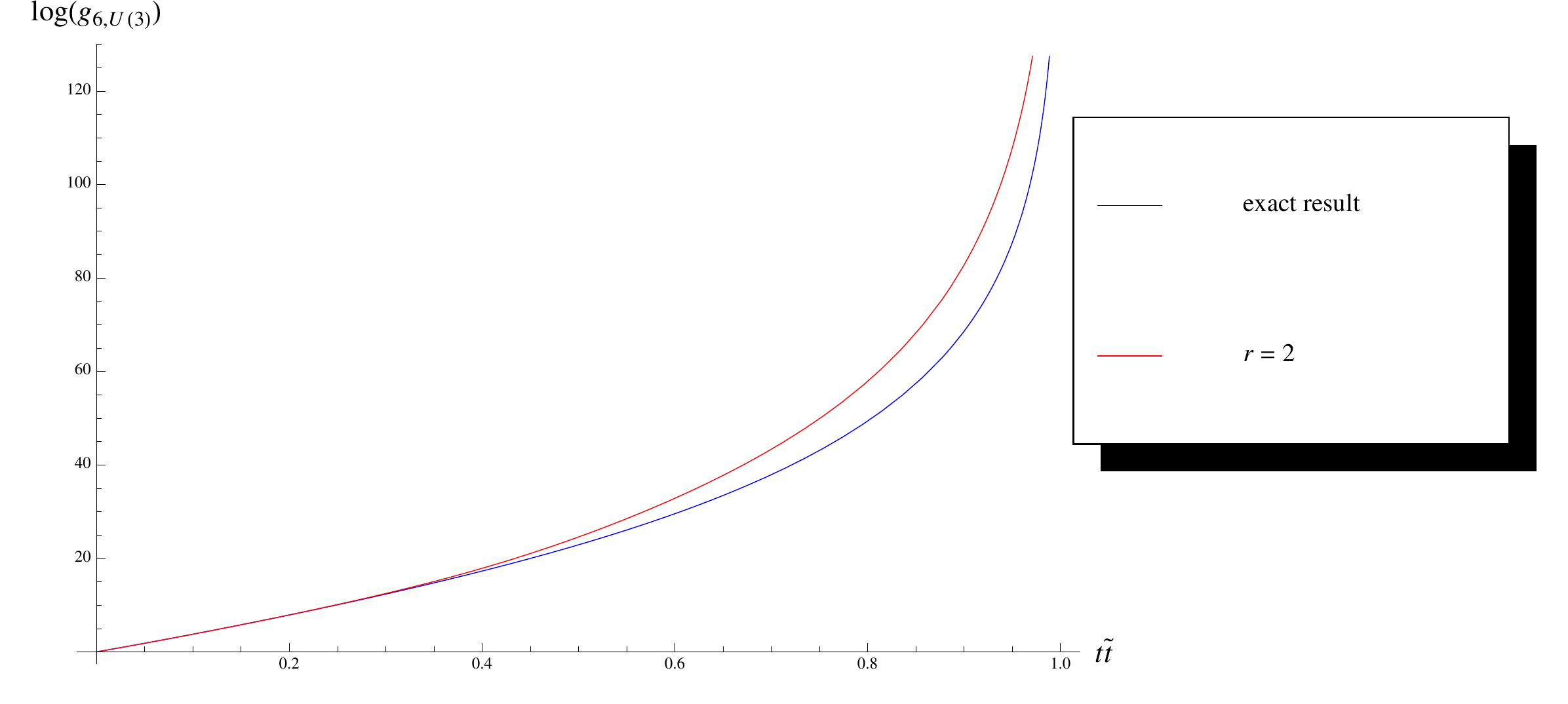}
\caption{The case of $N_c=3, N_f=6$: The graphs of the exact Hilbert series from \eref{nfncp3unref} and its asymptotic formula \eref{rncunc} (with $r=2$). As we expected from the second limiting case in \sref{sec:asymprncunc}, the asymptotic formula is in agreement with the exact result for $t \tti< 1/N_c = 1/3$.}
\label{fig:6U3}
\end{center}
\end{figure}

In \fref{fig:6U3}, we plot the graphs of $\log g_{6,U(3)}(t, \tti)$ given by \eref{nfncp3unref} and \eref{rncunc} (with $r=2$) against $t \tti$.  As we expected from the second limiting case in \sref{sec:asymprncunc}, the asymptotic formula should give a good approximation when $t \tti = O(N_c^{-1})$.  Indeed, as one can see from the graph, the asymptotic result is in agreement with the exact result for $t \tti< 1/N_c = 1/3$.

%The agreement between the exact result and the asymptotic formulae in \fref{fig:ncp2unc}, \fref{fig:ncp3unc} and \fref{fig:6U3} provide a non-trivial consistency checks for the formulae we have derived.

\section{$SU(N_c)$ SQCD with $N_f$ flavours}
Consider $\CN=1$ supersymmetric $SU(N_c)$ gauge theory in four dimensions with $N_f$ quark $Q^i_a$ and $N_f$ anti-quarks $\tQ^a_i$, where $a = 1, \ldots, N_c$ and $i = 1, \ldots, N_f$.  The superpotential of this theory is zero: $W=0$.  The information about the gauge and global symmetries as well as how the matters transform under such symmetries is collected in \tref{thisistabletwo} .

\begin{table}[htdp]
\begin{center}
{\small
\begin{tabular}{|c||c|cccccc|}
\hline
& Gauge symmetry & & & Global symmetry & & & \\
& $SU(N_c)$ & $SU(N_f)_1$ & $SU(N_f)_2$ & $U(1)_B$ & $U(1)_R$ & $U(1)_Q$ & $U(1)_{\widetilde{Q} }$\\
\hline \hline
$Q^i_a$ & $[0,\ldots,0,1]$ & $[1,0,\ldots,0]$ & $[0,\ldots,0]$ & 1 & $\frac{N_f-N_c}{N_f}$ & 1 & 0 \\
$\widetilde{Q}^a_i$ & $[1,0,\ldots,0]$ & $[0,\ldots,0]$ & $[0,\ldots,0,1]$ & $-1$ & $\frac{N_f-N_c}{N_f}$ & 0 & $-1$ \\
\hline
\end{tabular}}
\end{center}
\centering \includegraphics[height=1.8cm]{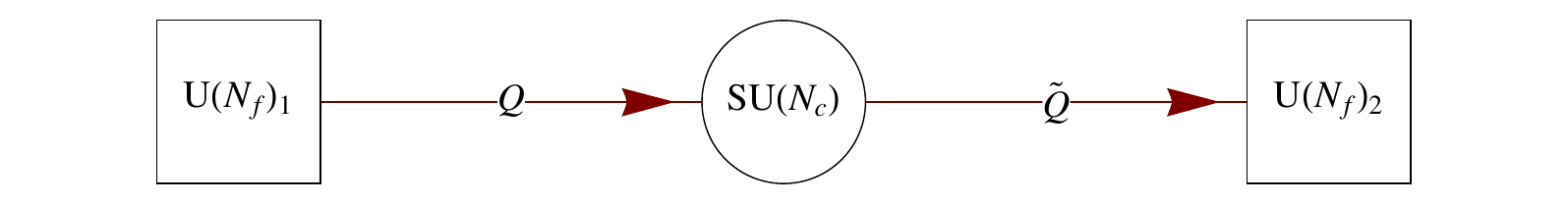}
\caption{The gauge and global symmetries of SQCD and the quantum numbers of the chiral supermultiplets. The quarks are $Q^i_a$ while the antiquarks are $\widetilde{Q}^a_i$. We also draw it as a quiver theory. The circular node represents the $U(N_c)$ gauge symmetry while the two square nodes represent global $U(N_f)_1$ and $U(N_f)_2$ symmetries. Each square node gives rise to a baryonic $U(1)$ global symmetry, one of which is redundant. We thus have $U(1)_{Q, \tilde{Q}}$ that combine into the non-anomalous $U(1)_B$ (sum) and anomalous $U(1)_A$ (difference).}
\label{thisistabletwo}
\end{table}

SQCD with the $SU(N_c)$ gauge group and $N_f$ flavours is studied and discussed in \cite{Seiberg:1994bz, Intriligator:1994sm, Seiberg:1994pq, argyres, terning, insei:07, insei:95, Gray:2008yu}.  Various geometrical aspects of the moduli space are examined in \cite{Gray:2008yu}, where the Hilbert series are also computed and studied in details.  In the following subsection, we present a method for computing the Hilbert series and discuss how to rewrite them in terms of Toeplitz  determinants.  We compare the method used in this paper with the one used in \cite{Gray:2008yu} in \sref{sec:difference}.

\subsection{The computations of Hilbert series} \label{sec:sunctpdet}
The Hilbert series for $SU(N_c)$ SQCD with $N_f$ flavours can be computed in a similar way to that for $U(N_c)$ SQCD with $N_f$ flavours.  The first step is to construct the Hilbert series of the space of symmetric functions of quarks $Q$ and antiquarks $\tQ$. This is given by
\bea
 \PE \left[ t [1,0,\ldots,0]_x \sum_{a=1}^{N_c} z_a^{-1} + \tti [0,0,\ldots,1]_y \sum_{a=1}^{N_c} z_a\right]~,
\eea
where for the $SU(N_c)$ gauge group, one needs to impose the condition that 
\bea z_1 z_2 \ldots z_{N_c}=1~.\eea
The second step is to integrate over the gauge group $SU(N_c)$ in order to obtain the Hilbert series which counts gauge invariant operators.  For this step, we need an appropriate Haar measure of $SU(N_c)$ for this problem.   Let us now discuss such a Haar measure.

\paragraph{The Haar measure.} The Haar measure of $SU(N_c)$ can be obtained from the Haar measure of $U(N_c)$ by restricting $z_1 \cdots z_{N_c}=1$.  This can be written as
\bea \label{haarsu1}
\int \ud \mu_{SU(N_c)}  &=& \frac{1}{N_c! (2 \pi i)^{N_c-1}} \oint_{|z_1| =1}  \frac{\ud z_1}{z_1} \cdots \oint_{|z_{N_c}| =1}  \frac{\ud z_{N_c}}{z_{N_c}}  |\Delta_{N_c}(z)|^2 \delta(z_1 \cdots z_{N_c} -1)~.\qquad
\eea
In Appendix \ref{sec:deltaintegral}, we show that the delta function in this multi-contour integral can be replaced by an infinite sum as follows:\footnote{We are very grateful to Harold Widom for pointing out this substitution to us and providing us with a note on analysis of Toeplitz determinant with a non-zero winding symbol.}
\bea
\delta(z_1 \cdots z_{N_c} -1) \quad \longrightarrow \quad \frac{1}{2 \pi i } \sum_{k = -\infty}^\infty z_1^k z_2^k \ldots z_{N_c}^k~.
\eea
Thus, the Haar measure of $SU(N_c)$ can be written as
\bea
\int \ud \mu_{SU(N_c)}  &=& \frac{1}{N_c! (2 \pi i)^{N_c}} \sum_{k = -\infty}^\infty \oint_{|z_1| =1}  \frac{\ud z_1}{z_1} \cdots \oint_{|z_{N_c}| =1}  \frac{\ud z_{N_c}}{z_{N_c}}  |\Delta_{N_c}(z)|^2 z_1^k \ldots z_{N_c}^k \nn \\
&=&  \sum_{k = -\infty}^\infty \int \ud \mu_{U(N_c)}  z_1^k \ldots z_{N_c}^k~.
\eea
This formula tells us that the Haar measure of the group $SU(N_c)$ can be written in terms of an infinite sum and the $U(N_c)$ Haar measure.

\paragraph{The Hilbert series.} Following from the above discussion, we see that the Hilbert series for $SU(N_c)$ SQCD with $N_f$ flavours is given by
\bea
&& g_{N_f, SU(N_c)} (t, \tti,x,y) = \int \ud \mu_{SU(N_c)} \PE \left[ t [1,0,\ldots,0]_x \sum_{a=1}^{N_c} z_a^{-1} + \tti [0,0,\ldots,1]_y \sum_{a=1}^{N_c} z_a\right] \nn \\
&=&  \sum_{k = -\infty}^{\infty} \int \ud \mu_{U(N_c)} z_1^k \ldots z_{N_c}^k \PE \left[ t [1,0,\ldots,0]_x \sum_{a=1}^{N_c} z_a^{-1} + \tti [0,0,\ldots,1]_y \sum_{a=1}^{N_c} z_a\right]~. \nn \\
\eea
Note that the expression in the square bracket is a Toeplitz determinant $D_{N_c}( \Phi)$ with the symbol $\Phi$ given by
\bea
\Phi(t, \tti, x, y, z; k) &=& z^k \PE \left[ t [1,0,\ldots,0]_x z^{-1} + \tti [0,0,\ldots,1]_y z\right] \nn \\
&=& z^k \phi(t, \tti, x, y, z) ~, \label{Phisymref}
\eea
where $\phi(t, \tti, x, y, z)$ is the symbol for the Toeplitz determinant for $U(N_c)$ SQCD, namely
\bea
\phi(t, \tti, x, y, z)  = \PE \left[ t [1,0,\ldots,0]_x z^{-1} + \tti [0,0,\ldots,1]_y z\right]~.
\eea

Note that, for $k \neq 0$, $\Phi$ has a non-zero winding number around the origin, \ie~
\bea
k = \frac{1}{2 \pi i} \oint \frac{\ud z}{z} \frac{\partial_z \Phi(t, \tti, x, y,z; k)}{\Phi(t, \tti, x, y,z; k)}~.
\eea
Note however that the method we have used in \sref{zerowindingtp} applies only to symbols with zero winding number. Therefore, such a method needs to be generalised in order to compute the Toeplitz determinant with the symbol $\Phi$.  In the next subsection, we discuss related theorems for the cases of non-zero winding numbers.  For now, we rewrite the Hilbert series of $SU(N_c)$ SQCD with $N_f$ flavours as
\bea
&& g_{N_f, SU(N_c)} (t, \tti,x,y) = D_{N_c} ( \phi) + \sum_{k=1}^\infty \left[ D_{N_c} (z^k \phi(t, \tti, x,y,z)) +D_{N_c} (z^{-k} \phi(t, \tti, x,y,z)) \right] \nn \\
&& \quad =g_{N_f, U(N_c)} (t, \tti,x,y) + \sum_{k=1}^\infty \left[ D_{N_c} (z^k \phi(t, \tti, x,y,z)) +D_{N_c} (z^{-k} \phi(t, \tti, x,y,z)) \right]~. \label{HSsumD}
\eea
This tells us that the Hilbert series of $SU(N_c)$ SQCD with $N_f$ flavours is simply the Hilbert series of $U(N_c)$ SQCD with $N_f$ flavours with correction terms coming from non-zero winding parts of the symbol.  Subsequently, we refer to the first, second and third terms in \eref{HSsumD} respectively as {\bf the zero winding part}, {\bf the positive winding part} and {\bf the negative winding part}. 

\paragraph{The unrefined Hilbert series.} Setting $x_1, \ldots, x_{N_c}$ and $y_1, \ldots, y_{N_c}$ to unity, we have the unrefined Hilbert series:
\bea
g_{N_f, SU(N_c)} (t, \tti)= \sum_{k = -\infty}^\infty \left[ \int \ud \mu_{U(N_c)} \prod_{a=1}^{N_c} \frac{z_a^k}{(1-t z_a)^{N_f} (1-\frac{\tti}{z_a})^{N_f}} \right]~. \label{SUNcSQCD}
\eea
Note that the expression in the square bracket is a Toeplitz determinant $D_{N_c}( \Phi)$ with the symbol $\Phi$ given by
\bea
\Phi(t, \tti, z; k) = z^k \phi(t, \tti, z) = \frac{z^k}{(1-\tti z)^{N_f} (1-\frac{t}{z})^{N_f}}~. \label{Phisym}
\eea

\subsubsection{Comparison with the method used in the `Aper\c{c}u paper'} \label{sec:difference}

We emphasise that the method of computation we use in this paper is significantly different from the one in the `Aper\c{c}u paper' \cite{Gray:2008yu}.  In this paper, we use coordinates $z_1, \ldots, z_{N_c}$ on the maximal torus of $U(N_c)$ and then impose the condition $\prod_{a=1}^{N_c} z_a =1$ using a delta function.  In this way, one can recast the Molien-Weyl formula into a Toeplitz determinant, which is a key tool for computations.   There are a number of techniques from the random matrix theory that can be used to evaluate such Toeplitz determinants both {\it exactly} and {\it asymptotically} for a large class of values of $N_c$ and $N_f$.  We present these techniques and apply them to our computations in the following subsections.

In \cite{Gray:2008yu}, on the other hand,  the integrand and the $SU(N_c)$ Haar measure are written in terms of $N_c-1$ coordinates on the maximal torus of $SU(N_c)$ without a delta function. In this way, one can compute Hilbert series case by case for a given $(N_f, N_c)$. These computations become very cumbersome when $N_c$ is large due to a large number of contour integrals.  However, based on case by case results, one can make conjectures about the general results, and indeed in \cite{Gray:2008yu} a number of these are stated as observations.  In the following subsections, a number of these observations can be proven (or at least can be checked in a non-trivial way) using Toeplitz determinants.  Moreover, we show that this technique can be used to compute a number of new exact Hilbert series and asymptotic formulae for large $N_f$ and $N_c$ -- many of which are too difficult or impossible to be derived using the method in \cite{Gray:2008yu}.

\subsection{The B\"ottcher-Widom and the Fisher-Hartwig theorems} \label{sec:BWFH}
In this section, we state two theorems which are very useful in computing the Toeplitz determinant with a symbol which has non-zero winding number around the origin.  In what follows, we still use the same definitions of the Toeplitz matrix, the Hankel matrix, $U$, $V$, $G(\phi)$ and $E(\phi)$ as in \S\S\ref{zerowindingtp} and \ref{BOSzego}.

\begin{theorem} {\bf (The B\"ottcher-Widom theorem. \cite{BottcherWidom}})  Let $\phi: \BT \rightarrow \BC -\{0\}$ be a continuous function such that $\phi(z) \neq 0$ for all $z \in \BT$ and $\phi$ has a zero winding number around the origin, \ie
\bea
0 = \frac{1}{2 \pi i} \oint_{|z|=1} \frac{\ud z}{z} \frac{\phi'(z)}{\phi(z)}~. 
\eea 
\bi
\item If $k>0$ and $T_{n+k}(\phi)$ is invertible, then both of the operators
\bea
\BU- H(b)  H(\widetilde{c}) Q_{n+k} \quad \text{and} \quad \BU- H(b) Q_n H(\widetilde{c}) Q_{k}
\eea
are invertible, and
\bea
D_{n}(z^{-k} \phi) =  (-1)^{n k} D_{n+k} (\phi) F_{n,k} (\phi)~,  \label{BW1}
\eea
where 
\bea
F_{n,k} ( \phi) = \det P_k (\BU - H(U) Q_n H(\tV)Q_k)^{-1} T(z^{-n} U) P_k~.
\eea

\item On the other hand, assuming $k> 0$, we have
\bea
D_n (z^k \phi) = (-1)^{n k} D_{n+k} (\phi) G(\phi)^{-k} G(U)^k F_{n,k} (\widetilde{\phi})~, \label{BW2}
\eea
where 
\bea
F_{n,k} (\widetilde{\phi}) = \det P_k T(z^{n} V) (\BU - Q_k H(U) Q_n H(\tV))^{-1}  P_k~.
\eea
\ei
\end{theorem}

The following theorem gives the asymptotic formula for $D_{n}(z^{-k} \phi) $ when $n$ is large.
\begin{theorem} \label{thm:FH} {\bf (The Fisher-Hartwig(-B\"ottcher-Silbermann) theorem. \cite{ref:FisherHartwig1,ref:FisherHartwig2, ref:BottcherSilbermann})} Assume $\phi$ to be as in the B\"ottcher-Widom Theorem.

\bi
\item If $k>0$, then 
\bea
F_{n,k} (\phi) \sim \det T_k (z^{-n} U)~.
\eea
It then follows from the B\"ottcher--Widom theorem and the strong Szeg\H{o} limit theorem that 
\bea
D_{n}(z^{-k} \phi) \sim  (-1)^{n k} G(\phi)^{n+k} E(\phi) \det T_k (z^{-n} U)~. \label{FH1}
\eea

\item On the other hand, assuming $k> 0$, we have
\bea
F_{n,k} (\widetilde{\phi}) \sim \det T_k (z^n V)~,
\eea
It then follows from the B\"ottcher--Widom theorem and the strong Szeg\H{o} limit theorem that  
\bea
D_{n}(z^{k} \phi) \sim  (-1)^{n k} G(\phi)^{n} E(\phi) G(U)^k  \det T_k (z^{n} V)~. \label{FH2}
\eea
\ei
\end{theorem}
Subsequently, we apply these theorems to compute Hilbert series of SQCD.

\subsubsection{Applications to the Hilbert series of SQCD}
Recall the discussion around \eref{Phisymref}. We take $\phi$ to be the symbol that we have used for $U(N_c)$ SQCD, \ie
\bea
\phi(t,\tti,x,y,z) =  \PE \left[ t [1,0,\ldots,0]_x z^{-1} + \tti [0,0,\ldots,1]_y z\right]~. \nn 
\eea
Then, the symbol $\Phi$ for the Toeplitz determinant of $SU(N_c)$ SQCD is 
\bea
\Phi(t, \tti, x, y, z; k) &=& z^k \PE \left[ t [1,0,\ldots,0]_x z^{-1} + \tti [0,0,\ldots,1]_y z\right] \nn \\
&=& z^k \phi(t, \tti, x, y, z)~. \nn
\eea
Hence, we can apply the B\"ottcher-Widom and Fisher-Hartwig theorems to compute the $N_c \times N_c$ Toeplitz determinant with the symbol $\Phi(t,\tti,x,y,z; k)$.  Thus, we take $n$ in these theorems to be the number of colours $N_c$.

\paragraph{Results for any $N_f$ and any $N_c$.} With the symbol $\Phi(t,\tti,x,y,z; k)$, we find that for any $N_c > 1,~N_f \geq 1$ and any $k \in \BZ$,
\bea
Q_{N_c} H(\widetilde{V}) Q_k &=& 0~, \nn\\
Q_k H(U) Q_{N_c} &=& 0~, \nn\\
G(U) &=& 1~. 
\eea
Let us now look at various cases of $N_f$ and $N_c$.

\subsubsection{The case of $N_f \leq N_c -1$} \label{sec:nfleqsuncm1}
For $N_f \leq N_c -1$, we find that, for all $k>0$,
\bea
P_k T(z^{-N_c} U) P_k = 0~,
\eea
and so it follows from the B\"ottcher-Widom theorem that
\bea
F_{N_c ,k} ( \phi) = 0~.
\eea
In other words, the Toeplitz determinants vanish for all negative winding number of the symbol $\Phi$, \ie~for all $k>0$,
\bea
D_{N_c} \left(\Phi(t, \tti, x, y, z; -k) \right) = D_{N_c} (z^{-k} \phi(t, \tti, x, y, z)) = 0~.
\eea

Similarly, for all $k>0$, we have
\bea
F_{N_c ,k} (\widetilde{ \phi}) = 0~,
\eea
and hence the Toeplitz determinants vanish for all positive winding number of the symbol $\Phi$, \ie~for all $k>0$,
\bea
D_{N_c} \left(\Phi(t, \tti, x, y, z; k) \right) = D_{N_c} (z^{k} \phi(t, \tti, x, y, z)) = 0~.
\eea

Therefore, it follows from \eref{HSsumD} that the Hilbert series for $SU(N_c)$ SQCD with $N_f \leq N_c-1$ flavours is exactly equal to the Hilbert series for $U(N_c)$ SQCD with $N_f \leq N_c-1$ flavours:
\bea \label{nfleqsuncm1}
&& g_{N_f \leq N_c-1} (t, \tti, x, y) = \PE \left[ [1,0, \ldots, 0;0, \ldots, 0,1]_{x;y} t \tti \right] \nn \\
&& = \frac{1}{1-(t \tti)^{N_f}} \sum_{n_1, \ldots, n_{N_f-1} = 0}^\infty [n_1, n_2, \ldots, n_{N_f-1};n_{N_f-1}, \ldots, n_2,n_1] (t \tti)^{\sum_{j=1}^{N_f-1} j n_j}~.\qquad
\eea
 The unrefined Hilbert series is then
\bea
g_{N_f \leq N_c-1} (t, \tti) = \frac{1}{(1-t \tti)^{N_f^2}}~.
\eea
Note that these Hilbert series are in agreement with (3.6) and (5.3) of \cite{Gray:2008yu}.

\paragraph{The moduli space for $N_f \leq N_c-1$.} As can be seen from the Hilbert series, the moduli space of $SU(N_c)$ SQCD with $N_f \leq N_c-1$ is freely generated by the meson transforming in the bi-fundamental representation $[1,0, \ldots, 0;0, \ldots, 0,1]$ of $SU(N_f) \times SU(N_f)$. 

Let us make a brief comment on quantum corrections. A non-perturbative Affleck--Dine--Seiberg (ADS) superpotential \cite{ADS, Seiberg:1994bz, argyres, terning} is dynamically generated.  This completely lifts the vacuum degeneracy and hence there is no supersymmetric vacuum.  While the Hilbert series of the classical moduli space does not have a physical meaning in the full quantum theory, it nevertheless contains information about the structure of gauge invariant operators for $N_f <N_c$.

\subsubsection{The case of $N_f = N_c$} \label{sec:nfeqsunc}
It follows from the B\"ottcher-Widom theorem that
\bea
F_{N_c ,k} ( \phi) = (-\tti)^{k N_c} ~, \qquad F_{N_c, k} (\widetilde{\phi}) = (-t)^{k N_c} ~.
\eea
Then, from \eref{BW1}, for $k >0$,
\bea
D_{N_c} (z^{-k} \phi) =  (-1)^{k N_c} D_{N_c+k} (\phi) (-\tti)^{k N_c} = D_{N_c+k} (\phi) \tti^{k N_c}~.
\eea
Recall that for $U(N_c+k)$ SQCD with $N_f = N_c$ flavours,
\bea
D_{N_c+k} (\phi) = g_{N_c, U(N_c+k)} (t, \tti, x, y) = \PE \left[ [1,0, \ldots, 0;0, \ldots, 0,1]_{x;y} t \tti \right]~.
\eea
Thus, we have
\bea
\sum_{k = 1}^\infty D_{N_c} (z^{-k} \phi) = \frac{\tti^{N_c}}{1-\tti^{N_c}} \PE \left[ [1,0, \ldots, 0;0, \ldots, 0,1]_{x;y} t \tti \right]~.
\eea
Similarly, using \eref{BW2}, we find that 
\bea
\sum_{k = 1}^\infty D_{N_c} (z^{k} \phi) = \frac{t^{N_c}}{1-t^{N_c}} \PE \left[ [1,0, \ldots, 0;0, \ldots, 0,1]_{x;y} t \tti \right]~.
\eea
The Hilbert series for $SU(N_c)$ SQCD with $N_f = N_c$ flavours is then given by \eref{HSsumD}:
\bea
&& g_{N_f,SU(N_c)} (t, \tti, x,y) = \left(1+ \frac{t^{N_c}}{1-t^{N_c}} +\frac{\tti^{N_c}}{1-\tti^{N_c}} \right) \PE \left[ [1,0, \ldots, 0;0, \ldots, 0,1]_{x;y} t \tti \right] \nn \\
&& \qquad =\frac{1- (t \tti)^{N_c}}{(1-t^{N_c}) (1-\tti^{N_c})}  \PE \left[ [1,0, \ldots, 0;0, \ldots, 0,1]_{x;y} t \tti \right] \nn \\
&& \qquad = \left(1- (t \tti)^{N_c} \right)  \PE \left[ [1,0, \ldots, 0;0, \ldots, 0,1]_{x;y} t \tti + t^{N_c} + \tti^{N_c} \right]~. \label{completeintsunc}
\eea
This can be written in terms of a sum of irreducible representations of $SU(N_f) \times SU(N_f)$ as 
\bea \label{nfeqsunc}
 g_{N_c , SU(N_c)}(t, \tilde{t},x,y) &=& \sum_{n_1, n_2, \ldots, n_{N_c-1}, \ell, m \geq 0} t^{\sum_{j=1}^{N_c-1} j n_j + \ell{N_c}}\: \tti^{ \sum_{j=1}^{N_c-1} j n_j + m{N_c}} \nn \\
&&  \qquad [n_1,n_2, \ldots, n_{N_c-1}~;~ n_{N_c-1} ,\ldots, n_2, n_1]_{x;y} ~.
\eea
This is in agreement with the general formula (5.2) of \cite{Gray:2008yu}.

Setting $x$'s and $y$'s to unity, we obtain the unrefined Hilbert series:
\bea
g_{N_c,SU(N_c)} (t, \tti, 1,1) &=&\frac{1- (t \tti)^{N_c}}{(1-t \tti)^{N_c^2}(1-t^{N_c}) (1-\tti^{N_c})} ~. \label{unrefsunfeqnc}
\eea
This is in agreement with (3.15) of \cite{Gray:2008yu}.  

\paragraph{The moduli space for $N_f = N_c$.}  
Indeed, this Hilbert series \eref{completeintsunc} tells us that the moduli space is a complete intersection (see, \eg, \cite{Gray:2008yu}).  The generators are mesons 
\bea M^i_j = Q^i_a \tQ^a_j~, \eea 
transforming in the bi-fundamental representation $[1,0, \ldots, 0;0, \ldots, 0,1]$ of $SU(N_f) \times SU(N_f)$, and baryons and anti-baryons
\bea B^{i_1 \ldots i_{N_c}} = \epsilon^{a_1 \ldots a_{N_c}}Q^{i_1}_{a_1} \ldots Q^{i_{N_c}}_{a_{N_c}}~, \qquad 
\tB_{i_1 \ldots i_{N_c}} = \epsilon_{a_1 \ldots a_{N_c}} \tQ_{i_1}^{a_1} \ldots \tQ_{i_{N_c}}^{a_{N_c}}~, \eea
each of which transforms as a singlet under $SU(N_f) \times SU(N_f)$.
There is one relation (at order  $(t \tti)^{N_c}$) between these generators, namely 
\bea
\det M - (*B)(*\tB) = 0~, \label{classrelnfnc}
\eea
where $*B = \frac{1}{N_c!} \epsilon_{i_1 \ldots i_{N_c}} B^{i_1 \ldots i_{N_c}}$ and similarly for $*\tB$.
Observe that the baryons, anti-baryons and their relation with mesons come from correction terms to the Hilbert series of $U(N_c)$ gauge theories with $N_f \leq N_c$ flavours.

Now let us make a brief comment on the quantum moduli space. It is still generated by $M$, $B$ and $\tB$. However, the classical relations \eref{classrelnfnc} is modified by a one instanton effect \cite{Seiberg:1994bz, Intriligator:1994sm, Seiberg:1994pq, argyres, terning, insei:07, insei:95}, and the quantum moduli space is described by the relation
\bea
\det M - (*B)(*\tB) = \Lambda^{2N_c}~,
\eea
where $\Lambda$ is the scale of the theory.
Although details of the relation are modified, the representations in which the generators and their relation transform under the global symmetry are unaffected. Thus, in spite of different geometrical properties between the classical and quantum moduli spaces, the Hilbert series is {\it not} corrected quantum mechanically.

\subsubsection{The case of $N_f = N_c+1$} \label{sec:nfsuncp1}
It follows from the B\"ottcher-Widom theorem that
\bea
F_{N_c ,k} ( \phi) &=& (-\tti)^{k N_c} \sum_{m=0}^k [0, \ldots, 0, 1_{m;L},0, \ldots,0~;~k-m,0, \ldots,0]_{x,y} (-t \tti)^m~, \nn \\
F_{N_c, k} (\widetilde{\phi}) &=& (-t)^{k N_c}  \sum_{m=0}^k [0,0, \ldots,k-m~;~0, \ldots, 0, 1_{m;R},0, \ldots,0]_{x,y} (-t \tti)^m~.
\eea
For $k > 0$, the Toeplitz determinant $D_{N_c +k} (\phi)$ is given by
\bea
D_{N_c +k} (\phi)= g_{N_c+1, U(N_c+k)} (t, \tti, x, y) = \PE \left[ [1,0, \ldots, 0;0, \ldots, 0,1]_{x;y} t \tti \right]~,
\eea
and the Toeplitz determinant $D_{N_c} (\phi)$ is given by
\bea
D_{N_c } (\phi)= g_{N_c+1, U(N_c)} (t, \tti, x, y) = (1- (t \tti)^{N_c+1} )\PE \left[ [1,0, \ldots, 0;0, \ldots, 0,1]_{x;y} t \tti \right]~.
\eea
Thus, from \eref{HSsumD}, we find that the Hilbert series of $SU(N_c)$ SQCD with $N_f = N_c+1$ flavours can be written as
{\small
\bea
&&  g_{N_c+1, SU(N_c)} (t, \tti, x, y) = \Bigg[1- (t \tti)^{N_c+1}  + \sum_{k=1}^\infty \tti^{k N_c} \sum_{m=0}^k [0, \ldots, 0, 1_{m;L},0, \ldots,0~;~k-m,0, \ldots,0]_{x,y} (-t \tti)^m \nn \\
&& + \sum_{k=1}^\infty t^{k N_c} \sum_{m=0}^k [0, \ldots,0,k-m~;~0, \ldots, 0, 1_{m;R},0, \ldots,0]_{x,y} (-t \tti)^m\Bigg]  \PE \left[ [1,0, \ldots, 0;0, \ldots, 0,1]_{x;y} t \tti \right]~.\nn \\
\eea
}
This can be rewritten in terms of a sum of irreducible representations of $SU(N_f) \times SU(N_f)$ as
\bea
g_{N_c+1, SU(N_c)}(t, \tilde{t},x,y) &=& \sum_{n_1, n_2, \ldots, n_{N_c-1}, \ell, m \geq 0} t^{\sum_{j=1}^{N_c-1} j n_j + \ell{N_c}}\: \tti^{ \sum_{j=1}^{N_c-1} j n_j + m{N_c}} \times \nn \\
&&  \qquad [n_1,n_2, \ldots, n_{N_c-1}, \ell~;~  m, n_{N_c-1} ,\ldots, n_2, n_1]_{x;y} ~. \label{refncp1sunc}
\eea
This is in agreement with the general formula (5.26) of \cite{Gray:2008yu}.

\paragraph{The unrefined Hilbert series.} Setting $x$'s and $y$'s to unity, we find that
\bea
F_{N_c ,k} ( \phi) &=& (-\tti)^{k N_c} \sum_{m=0}^k {N_f \choose m }{N_f+k-m-1 \choose k-m } (-t \tti)^m \nn \\
&=& (-\tti)^{k N_c} {N_f+k-1 \choose k} {}_2F_1(-k,-N_f;1-k-N_f; t \tti)~, \nn \\
F_{N_c ,k} ( \widetilde{\phi}) &=& (-t)^{k N_c} \sum_{m=0}^k {N_f \choose m }{N_f+k-m-1 \choose k-m } (-t \tti)^m \nn \\
&=& (-t)^{k N_c} {N_f+k-1 \choose k} {}_2F_1(-k,-N_f;1-k-N_f; t \tti)~.
\eea
Then, we find that the negative winding part is
\bea
&& \sum_{k=1}^\infty D_{N_c} (z^{-k} \phi) = \frac{1}{(1- t \tti)^{N_f^2}} \sum_{k=1}^\infty \tti^{k N_c} {N_f+k-1 \choose k} {}_2F_1(-k,-N_f;1-k-N_f; t \tti) \nn\\
&& =  \frac{1}{(1- t \tti)^{N_f^2-1}(1- \tti^{N_c})^{N_f}} \sum_{m= 0}^{N_c} \sum_{n=0}^{N_f} (-1)^{m+n} {N_f \choose m+n+1} \tti^{(m+1) N_c +n N_f} t^{n}~, \nn \\
\eea
whereas the positive winding part is
\bea
 \sum_{k=1}^\infty D_{N_c} (z^{k} \phi)  &=&  \frac{1}{(1- t \tti)^{N_f^2-1} (1- t^{N_c})^{N_f}}  \times \nn \\
 &&  \sum_{m= 0}^{N_c} \sum_{n=0}^{N_f} (-1)^{m+n} {N_f \choose m+n+1} t^{(m+1) N_c +n N_f} \tti^{n}~, 
\eea
and from \eref{unrefnfncp1}, we find that the zero winding part is
\bea
D_{N_c} ( \phi)= \frac{1- (t \tti)^{N_f}}{ (1- t \tti)^{N_f^2}}~.
\eea
Thus, from \eref{HSsumD}, we obtain the unrefined Hilbert series for $SU(N_c)$ with $N_f=N_c+1$ flavours as
\bea \label{exactncp1sunc}
&& g_{N_c+1, SU(N_c)} (t, \tilde{t}) 
%=  \frac{1}{(1- t \tti)^{N_f^2}}   \Bigg[ 1- (t \tti)^{N_f} +\nn \\
%&&  (1- t \tti)  \sum_{m= 0}^{N_c} \sum_{n=0}^{N_f} (-1)^{N_c+m+n+1} {N_f \choose m+n+1} \left(\frac{t^{m N_c +n N_f} \tti^{n} }{(1- t^{N_c})^{N_f}}+ \frac{\tti^{m N_c +n N_f} t^{n}}{(1- \tti^{N_c})^{N_f}} \right) \Bigg]  \nn \\
=  \frac{1}{(1- t \tti)^{(N_c+1)^2}}   \Bigg[ 1- (t \tti)^{N_c+1} + \nn \\
&&  (1- t \tti)  \sum_{m= 0}^{N_c} \sum_{n=0}^{N_c+1} (-1)^{m+n} {N_c+1 \choose m+n+1} \left(\frac{t^{(m+n+1) N_c} }{(1- t^{N_c})^{N_c+1}}+ \frac{\tti^{(m+n+1) N_c }}{(1- \tti^{N_c})^{N_c+1}} \right) (t \tti)^n \Bigg]~. \nn \\
\eea

\paragraph{Examples.} For simplicity, we set $t = \tti$. From \eref{exactncp1sunc}, we have
\bea
g_{3,SU(2)} (t) &=& \frac{1+6 t^2+6 t^4+t^6}{\left(1-t^2\right)^9}~, \nn \\
g_{4, SU(3)} (t) &=& \frac{1+4 t^2+4 t^3+10 t^4+8 t^5+14 t^6+8 t^7+10 t^8+4 t^9+4 t^{10}+t^{12}}{(1-t^2)^{12} (1-t^3)^4}~, \nn \\
g_{5, SU(4)} (t) &=& \frac{1+5 t^2+20 t^4+50 t^6+85 t^8+100 t^{10}+85 t^{12}+50 t^{14}+20 t^{16}+5 t^{18}+t^{20}}{\left(1-t^2\right)^{25} \left(1+t^2\right)^5}~.\nn 
\eea
Note that the first two results are in agreement with those in \cite{Gray:2008yu}.  These examples provide a consistency check of the formula \eref{exactncp1sunc}. We emphasise here that with the knowledge of Toeplitz matrices and their determinants, one can obtain exact results without encountering difficulties from computing a large number of contour integrals.  

\paragraph{The moduli space for $N_f > N_c$.} For completeness of the paper, let us briefly summarise the information about the moduli space for $N_f > N_c$ (see also \cite{Seiberg:1994bz, Intriligator:1994sm, Seiberg:1994pq, argyres, terning, insei:07, insei:95}). This information is contained in the Hilbert series and can be extracted using the plethystic logarithm.  Since this has already been shown in \cite{Gray:2008yu}, we simply state the results here.
The generators are mesons 
\bea M^i_j = Q^i_a \tQ^a_j~, \eea 
transforming in the bi-fundamental representation $[1,0, \ldots, 0;0, \ldots, 0,1]$ of $SU(N_f) \times SU(N_f)$, and baryons and anti-baryons
\bea B^{i_1 \ldots i_{N_c}} = \epsilon^{a_1 \ldots a_{N_c}}Q^{i_1}_{a_1} \ldots Q^{i_{N_c}}_{a_{N_c}}~, \qquad 
\tB_{i_1 \ldots i_{N_c}} = \epsilon_{a_1 \ldots a_{N_c}} \tQ_{i_1}^{a_1} \ldots \tQ_{i_{N_c}}^{a_{N_c}}~; \eea
transforming respectively in the representation 
\bea
[0,\ldots,0,1_{N_c;L},0, \ldots,0;0, \ldots,0] \quad \text{and} \quad [0, \ldots,0; 0,\ldots,0,1_{N_c;R},0, \ldots,0]
\eea
of $SU(N_f) \times SU(N_f)$.  They are subject to the relations:
\bea
(*B)\tB = *(M^{N_c})~,\quad M \cdot *B =0~, \quad M \cdot *\tB=0~.
\eea
where $(*B)_{i_{N_c+1} \ldots i_{N_f}} = \frac{1}{N_c!} \epsilon_{i_1 \ldots i_{N_f}} B^{i_1 \ldots i_{N_c}}$ and a `$\cdot$' denotes a contraction of an upper with a lower flavour index.  These relations transform respectively in the representations
\bea
&& [0,...,0,1_{N_c;L},0,...,0;0,...,0,1_{N_c;R},0,...,0]~, \nn \\
&& [0,...,0,1_{(N_c+1);L},0,...,0;0,...,0,1]~, [1,0,...,0;0,...,0,1_{(N_c+1);R},0,...,0]~.
\eea
Note that, in this case, the quantum moduli space coincides with the classical moduli space \cite{Seiberg:1994bz, insei:07}. Thus, the Hilbert series of the classical moduli space for $N_f > N_c$ is also valid for the quantum moduli space of the theory.

\subsection{An exact refined Hilbert series for any $N_f$ and $N_c$}
From \eref{nfleqsuncm1}, \eref{nfeqsunc} and \eref{refncp1sunc}, one can see that the results we have obtained so far are in agreement with the general formula (5.26) of \cite{Gray:2008yu} which gives the Hilbert series for any $N_f$ and any $N_c$:
\bea \label{suncexactref}
&& g_{N_f , SU(N_c)}(t, \tilde{t}) = \sum_{n_1, n_2, \ldots, n_{N_c-1}, \ell, m \geq 0} t^{\sum_{j=1}^{N_c-1} j n_j + \ell{N_c}}\: \tti^{ \sum_{j=1}^{N_c-1} j n_j + m{N_c}} \nn \\
&&  \qquad [n_1,n_2, \ldots, n_{N_c-1}, \ell_{N_c;L},0, \ldots,0; 0,\ldots, 0 , m_{N_c;R}, n_{N_c-1} ,\ldots, n_2, n_1] ~.
\eea
%where $k=N_c-1$, $a = \ell{N_c}+  \sum_{j=1}^k j n_j$, $b = m{N_c}+ \sum_{j=1}^k j n_j$

So far we have discuss a number of exact results for $SU(N_c)$ SQCD with $N_f$ flavours.  In the next subsection, we study asymptotics of the Hilbert series for large $N_f$ and $N_c$.

\subsection{Asymptotics for large $N_f$ and $N_c$} \label{sec:asympsu}

In the limit of large $N_f$ and $N_c$, it is convenient to obtain the Hilbert series using the Fisher-Hartwig Theorem \ref{thm:FH}. 

\paragraph{The negative winding part.} Let us first consider the negative winding part in \eref{HSsumD}.   We would like to evaluate the sum $\sum_{k=1}^\infty D_{N_c} (z^{-k} \phi)$ using \eref{FH1}.  (Recall that $G(\phi) = 1$ and $G(U) = 1$.) We claim that the leading contribution to this sum comes from the term with $k=1$, namely
\bea
D_{N_c} (z^{-1} \phi) = (-1)^{N_c} E( \phi) \det T_1 (z^{-N_c}U) = (-1)^{N_c} E( \phi) T_1 (z^{-N_c}U)~.  \nn
\eea
This is because for $N_f, N_c >>1$, the leading behaviour of $\det T_k (z^{-N_c}U)$ is $\tti^{kN_c}$; therefore $D_{N_c} (z^{-k} \phi)$ is of order $\tti^{kN_c}E(\phi)$, and hence the other terms in the sum $\sum_{k=1}^\infty D_{N_c} (z^{-k} \phi)$ can be neglected in comparison with $D_{N_c} (z^{-1} \phi)$.  It follows immediately from the definition \eref{def:Toeplitzmat} of the Toeplitz matrix that
\bea
T_1 (z^{-N_c}U) = \frac{1}{2 \pi i} \oint_{|z|=1} \frac{\ud z}{z} z^{-N_c} U = U_{N_c}~.
%&=& (-\tti)^{N_c}   \sum_{m=0}^{\delta} [m,0, \ldots,0~;~0, \ldots, 0, 1_{(N_c+m);R},0, \ldots,0]_{x; y} (-t \tti)^{m} \nn\\
%&=& U_{N_c}~,
\eea
where we have used \eref{UVdef} to establish the last equality.  Therefore, using \eref{FH1} and recalling from \eref{Ephi} that
\bea
E(\phi) = \PE \left[ [1,0, \ldots, 0;0, \ldots, 0,1]_{x;y} t \tti \right]~, \nn 
\eea
we obtain
\bea
\sum_{k=1}^\infty D_{N_c} (z^{-k} \phi) &\sim& D_{N_c} (z^{-1} \phi) \nn \\
&=&  (-1)^{N_c} E( \phi) T_1 (z^{-N_c}U) \nn \\
&=& (-1)^{N_c} U_{N_c}  \PE \left[ [1,0, \ldots, 0;0, \ldots, 0,1]_{x;y} t \tti \right]~,
\eea
where we have neglected terms of order $O(\tti^{2N_c}) E(\phi)$ and smaller.

\paragraph{The positive winding part.} A similar argument can be applied to compute the positive winding part of \eref{HSsumD}.  From \eref{def:Toeplitzmat} and \eref{UVdef}, we find that 
\bea
\det T_1 (z^{N_c} V) = T_1 (z^{N_c} V)  = \frac{1}{2 \pi i} \oint_{|z|=1} \frac{\ud z}{z} z^{N_c} V= V_{-N_c}~.
\eea
%On the other hand, $\det T_k (z^{N_c}U)$ is of order $t^{kN_c}$, and for $k=1$, we have
%\bea
%\det T_1 (z^{N_c}U) &=& (-t)^{N_c}   \sum_{m=0}^{\delta} [0, \ldots, 0, 1_{(N_c+m);L},0, \ldots,0~;~ 0, \ldots,0,m]_{x; y} (-t \tti)^{m} \nn\\
%&=& V_{N_c}~,
%\eea
%where we have used \eref{UVref} in the last equality.  
Therefore, from \eref{FH2}, we obtain
\bea
\sum_{k=1}^\infty D_{N_c} (z^{k} \phi) &\sim& D_{N_c} (z \phi) \nn \\
&=&  (-1)^{N_c} E( \phi) T_1 (z^{N_c}V) \nn \\
&=& (-1)^{N_c} V_{-N_c}  \PE \left[ [1,0, \ldots, 0;0, \ldots, 0,1]_{x;y} t \tti \right]~,
\eea
where we have neglected terms of order $O(t^{2N_c}) E(\phi)$ and smaller.

\paragraph{The zero winding part.} The zero winding part $D_{N_c} ( \phi)$ is discussed in \sref{sec:variousasympunc}. 
Recall that there are 3 limiting cases of our interest:
\bi
\item The difference $\Delta : = N_f -(N_c+1) \geq 0$ is finite when $N_f, N_c \rightarrow \infty$.
\item The ratio $r := N_f/N_c  = 1+ o(1)$ as $N_c \rightarrow \infty$.  (This is case 1 in \sref{sec:asymprncunc}.)
\item The ratio $r \geq 1$ is finite and $t \tti = o(1)$ as $N_c \rightarrow \infty$.  (This is case 2 in \sref{sec:asymprncunc}.)
\ei
In these limiting cases, the following approximation is valid:
\bea
D_{N_c} ( \phi) \sim (1- U_{N_c + 1} V_{-(N_c + 1)}) \PE \left[ [1,0, \ldots, 0;0, \ldots, 0,1]_{x;y} t \tti \right]~.
\eea

\paragraph{The asymptotic formula.} Thus, from \eref{HSsumD}, we find that
\bea
g_{N_f, SU(N_c)} (t, \tti, x,y) &\sim& \Big(1 +(-1)^{N_c} U_{N_c}+(-1)^{N_c} V_{-N_c} - U_{N_c + 1} V_{-(N_c + 1)}\Big) \times \nn \\
&&  \PE \left[ [1,0, \ldots, 0;0, \ldots, 0,1]_{x;y} t \tti \right]~. \label{nfsuncasymp}
\eea

\subsubsection{Asymptotics for $N_f, N_c >>1$ with a fixed difference $N_f -N_c \geq0$} \label{sec:asympfixeddiffsu}
Now let us focus on the limit $N_f, N_c >>1$ with a fixed difference
\bea \delta := N_f -N_c \geq 0~. \eea
From \eref{un} and \eref{vn}, we have the following asymptotic formulae:
\bea \label{UVncasymp}
U_{N_c} &=& (-\tti)^{N_c}  {N_f \choose N_c} {}_2F_1 ( -\delta, N_f; N_f - \delta +1; t \tti)  \nn \\
&\sim& (-\tti)^{N_c}  {N_f \choose N_c} {}_2F_1 ( -\delta, N_f; N_f; t \tti)   = {N_f \choose N_c} (1- t \tti)^{\delta} (-\tti)^{N_c} ~, \nn \\
V_{-N_c} &\sim& {N_f \choose N_c} (1- t \tti)^{\delta} (-t)^{N_c} ~. 
\eea
Substituting \eref{UVncasymp} and \eref{apptrknc} into \eref{nfsuncasymp}, we find that the asymptotic formula for the unrefined Hilbert series is 
\bea
&& g_{N_c + \delta, SU(N_c)} (t, \tti) \sim \frac{1}{(1- t \tti)^{(N_c+\delta)^2}} \times \nn \\
&& \qquad  \left[1 +{N_c+ \delta \choose N_c} (1- t \tti)^{\delta} (t^{N_c} + \tti^{N_c}) - {N_c +\delta \choose N_c+1}^2 (1- t \tti)^{2(\delta+1)} (t \tti)^{N_c+1} \right]~.\qquad \label{asympdeltasu}
\eea

\paragraph{Special case of $N_f = N_c$.} In this special case, the formula \eref{asympdeltasu} reduces to
\bea \label{asympsuNfeqNc}
g_{N_c SU(N_c)} (t, \tti, x,y) \sim \frac{1 +t^{N_c} + \tti^{N_c}}{(1- t \tti)^{N_c^2}}~.
\eea
It is worthwhile comparing this with \eref{unrefsunfeqnc}. We see that in the limit of large $N_c$,
\bea
\frac{1-(t \tti)^{N_c}}{(1-t^{N_c})(1-\tti^{N_c})} \sim 1 +t^{N_c} + \tti^{N_c} ~.
\eea
Substituting this into \eref{unrefsunfeqnc}, we recover \eref{asympsuNfeqNc} as expected.

We provide other consistency checks for \eref{asympdeltasu} in \sref{sec:conschecksu}.

\subsubsection{Asymptotics for $N_f, N_c>>1$ with a fixed ratio $ N_f/N_c \geq 1$} \label{sec:asympfixedratio}
Let $r$ be the the ratio $N_f/ N_c$ between $N_f$ and $N_c$ and assume that $r \geq 1$.
We apply \eref{nfsuncasymp} to find the asymptotic formula. 

Recall from \eref{asympUVunc} and \eref{stirling1} that 
\bea
U_{N_c+1} V_{-(N_c+1)} &\sim& {r N_c \choose N_c+1}^2  \left[ {}_2F_1 (-(r-1)N_c+1, r N_c; N_c+2; t \tti) \right]^2 (t \tti)^{N_c+1}~. \quad
%&\sim& \frac{1}{2 \pi N_c} \left[ \frac{r^{2N_c r+1} }{(r-1)^{2N_c(r-1)-1}} \right] \left[ \CF(N_c, r, t \tti) \right]^2 (t \tti)^{N_c+1}~,
\eea
where $\CF(N_c, r, t \tti)$ is defined as
\bea
\CF(N_c, r, t \tti) := {}_2F_1 ( -(r-1)N_c +1 , r N_c; N_c +2; t \tti)~. \nn
\eea
Now let us compute $(-1)^{N_c} U_{N_c}+(-1)^{N_c} V_{-N_c}$. From \eref{un} and \eref{vn}, we can write
\bea
U_{N_c} =  (-\tti)^{N_c}  {r N_c \choose N_c} {}_2F_1 ( -(r-1)N_c , r N_c; N_c +1; t \tti)~, \nn \\
V_{-N_c} =  (-t)^{N_c}  {r N_c \choose N_c} {}_2F_1 ( -(r-1)N_c , r N_c; N_c +1; t \tti)~.
\eea
%Using the Stirling formula (c.f. the derivation of \eref{stirling1}), we find that
%\bea
%{rN_c \choose N_c} 
%\sim \frac{1}{\sqrt{2 \pi N_c}} \left[ \frac{r^{r N_c +\frac{1}{2}} }{(r-1)^{(r-1)N_c +\frac{1}{2}}} \right]~.
%\eea
Writing 
\bea
\widehat{\CF}(N_c, r, t \tti) := {}_2F_1 ( -(r-1)N_c  , r N_c; N_c +1; t \tti)~,
\eea
we have 
\bea
(-1)^{N_c} U_{N_c}+(-1)^{N_c} V_{-N_c} \sim {r N_c \choose N_c} \widehat{\CF}(N_c, r, t \tti) \left( t^{N_c} + \tti^{N_c} \right)~. \quad
\eea
Thus, from \eref{nfsuncasymp}, we arrive at the asymptotic formula
\bea \label{asymprncsunc}
 g_{rN_c, SU(N_c)} (t ,\tti) &\sim& \frac{1}{(1-t \tti)^{r^2 N_c^2}} \Bigg[ 1+ {r N_c \choose N_c} \widehat{\CF}(N_c, r, t \tti) \left( t^{N_c} + \tti^{N_c} \right) \nn \\
&& \qquad - {r N_c \choose N_c+1}^2 \left[ \CF(N_c, r, t \tti) \right]^2 (t \tti)^{N_c+1} \Bigg]~. 
\eea

One interesting application of this asymptotic formula is that one can use it to study the moduli space in the conformal window, namely for $\frac{3}{2}N_c < N_f < 3N_c$, where the theory has a non-trivial IR fixed point. In the conformal window, the theory possesses a dual description, known as the {\it Seiberg duality} \cite{Seiberg:1994pq}. It is an interesting problem to check this duality using Hilbert series. So far R\"omelsberger \cite{Romelsberger:2005eg} has showed that the Hilbert series of $SU(2)$ SQCD with 3 flavours and its magnetic dual match.  Now that the Hilbert series for the theory in the conformal window is available in the limit of large $N_c$ and $N_f$, the checking should be possible for such an asymptotic limit.  Since this problem deserves investigation in its own right, we leave this to future work.

\subsubsection{Consistency checks} \label{sec:conschecksu}
In this subsection, we provide consistency checks for various asymptotic formulae we have derived so far.  

\begin{figure}[htbp]
\begin{center}
\includegraphics[height=6cm]{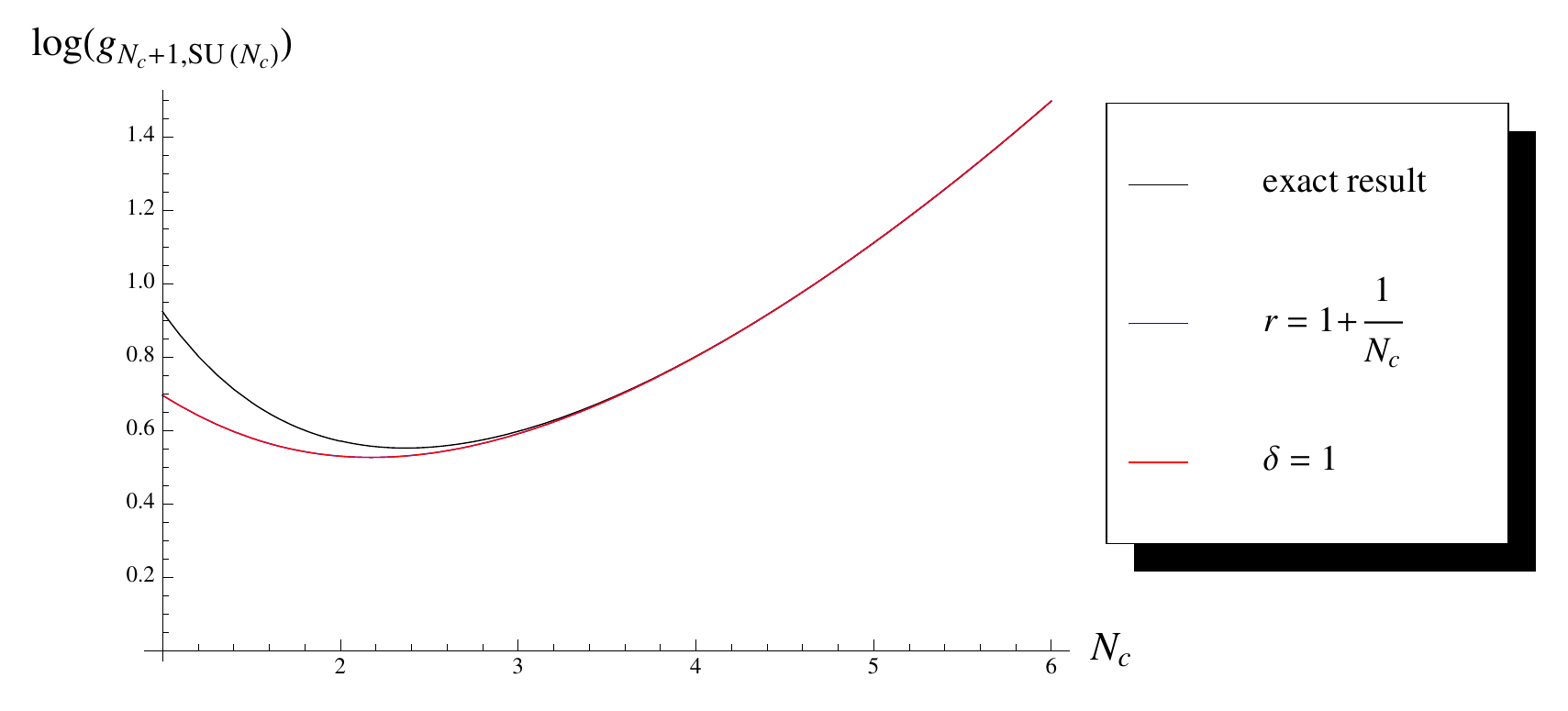}
\caption{The case of $N_f = N_c+1$: The graphs of $\log g_{N_c+1,SU(N_c)}(t, \tti)$ given by \eref{exactncp1sunc} with its asymptotic formulae \eref{asympdeltasu} and \eref{asymprncsunc}. We take $\delta =1$, $r=1+ 1/N_c$, $t = 0.3$ and $\tti = 0.1$.  Note that the graphs for both asymptotic formulae are on top of each other in this figure.}
\label{fig:ncp1sunc}
\end{center}
\end{figure}

In \fref{fig:ncp1sunc}, we plot the graphs of $\log g_{N_c+1,SU(N_c)}(t, \tti)$ given by \eref{exactncp1sunc}, \eref{asympdeltasu} and \eref{asymprncsunc} against $N_c$, with $\delta =1$, $r=1+ 1/N_c$, $t = 0.3$ and $\tti = 0.1$.  It can be see that the asymptotic formulae \eref{asympdeltasu} and \eref{asymprncsunc} approach the exact result \eref{exactncp1sunc} when $N_c$ is large.

\begin{figure}[htbp]
\begin{center}
\includegraphics[height=6cm]{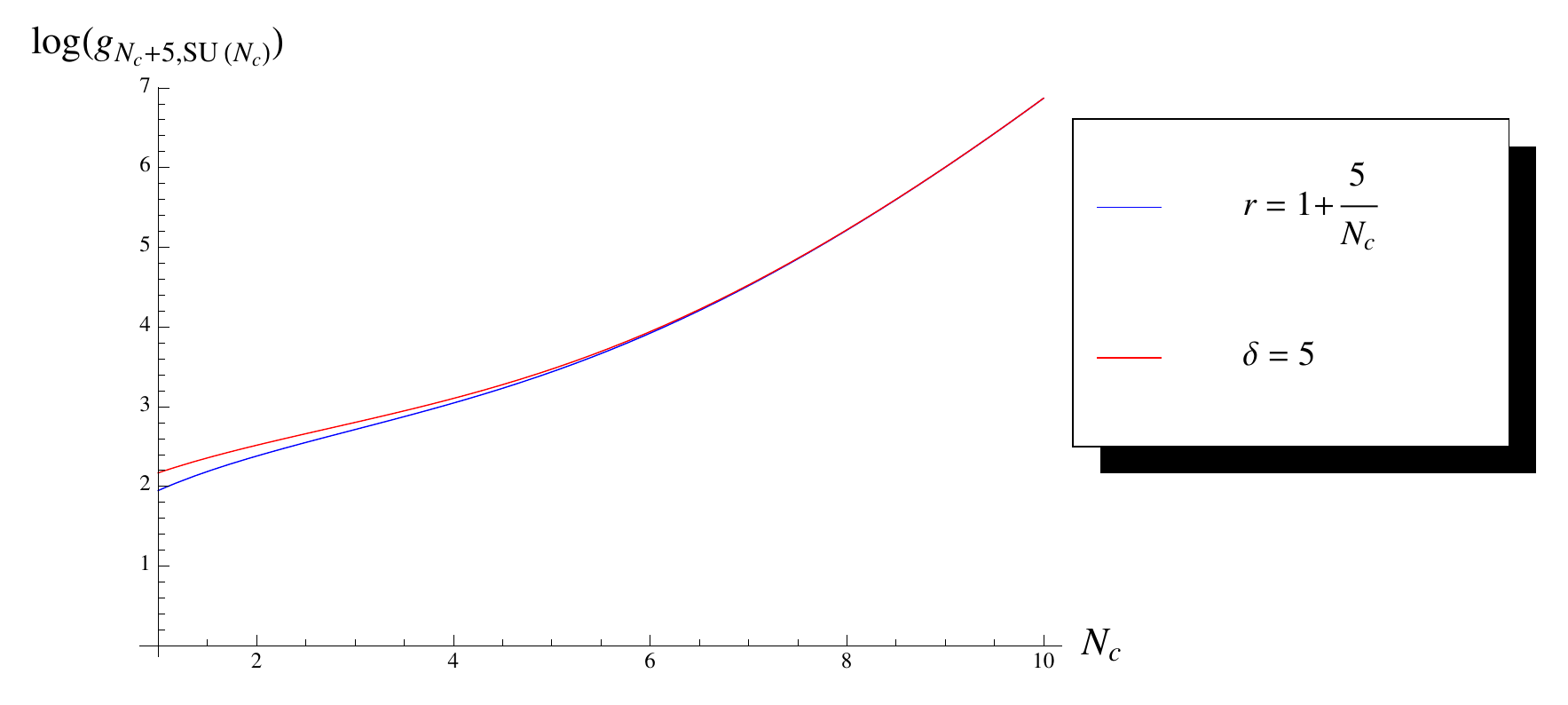}
\caption{The case of $N_f = N_c+5$: The graphs of $\log g_{N_c+5,SU(N_c)}(t, \tti)$ given by the asymptotic formulae \eref{asympdeltasu} and \eref{asymprncsunc}.  We take $\delta =5$, $r=1+ 5/N_c$, $t = 0.3$ and $\tti = 0.1$.}
\label{fig:ncp5sunc}
\end{center}
\end{figure}

In \fref{fig:ncp5sunc}, we plot the graphs of $\log g_{N_c+5,SU(N_c)}(t, \tti)$ given by the asymptotic formulae \eref{asympdeltasu} and \eref{asymprncsunc} against $N_c$, with $\delta =5$, $r=1+ 5/N_c$, $t = 0.3$ and $\tti = 0.1$.  It can be see that the asymptotic formula \eref{asympdeltasu} approaches the asymptotic formula \eref{asymprncsunc} for large $N_c$.

\begin{figure}[htbp]
\begin{center}
\includegraphics[height=6cm]{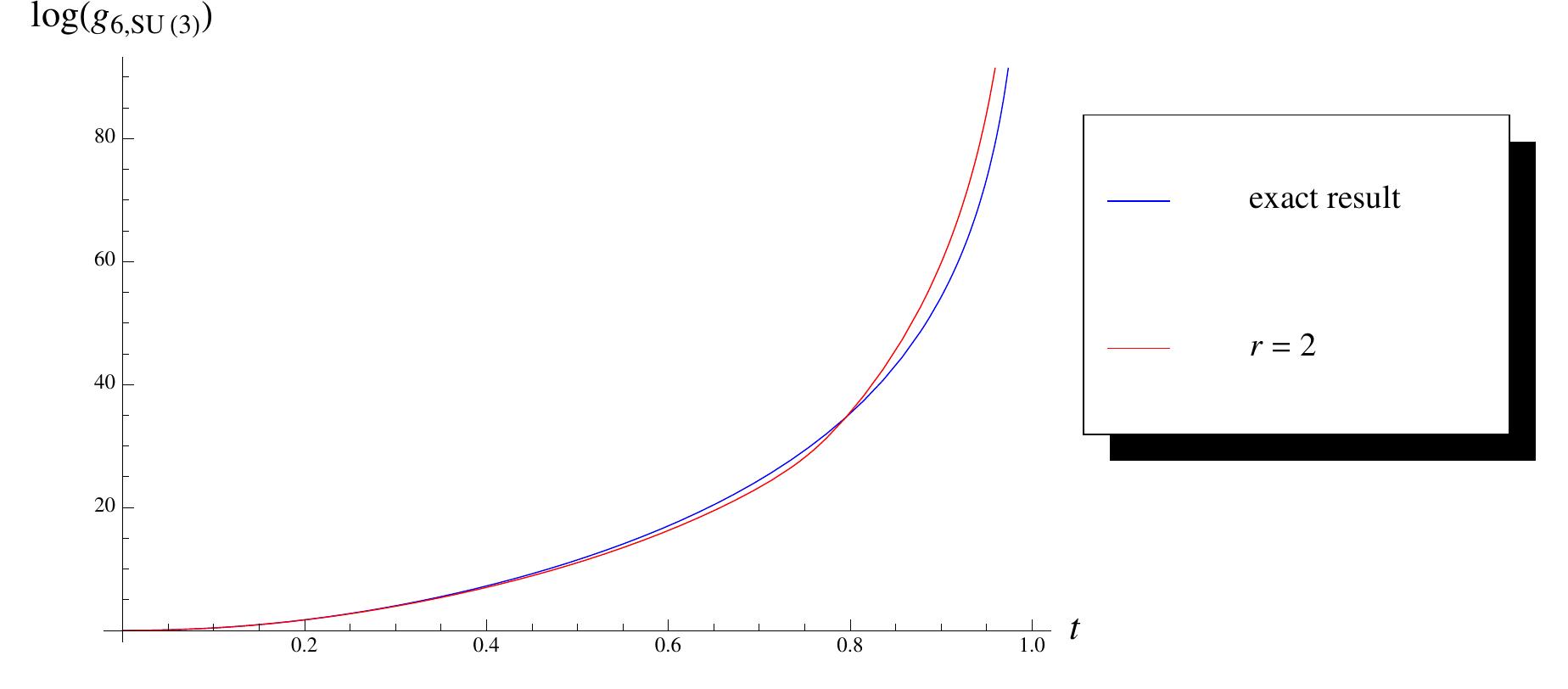}
\caption{The case of $N_c=3, N_f=6$ and $t = \tti$: The graphs of $\log g_{6,SU(3)}(t, t)$ from \eref{suncexactref}  and its asymptotic formula \eref{asymprncsunc} (with $r=2$ and $N_c=3$). As we expected from the second limiting case in \sref{sec:asymprncunc}, the asymptotic formula is in agreement with the exact result for $t < 1/\sqrt{N_c} = 1/\sqrt{3} \approx 0.58$.}
\label{fig:plot6su3}
\end{center}
\end{figure}

Let us now set $t = \tti$. In \fref{fig:plot6su3}, we plot the graphs of $\log g_{6,SU(3)}(t, t)$ given by \eref{suncexactref} and \eref{asymprncsunc} (with $r=2$ and $N_c=3$) against $t$.  As we expected from the second limiting case in \sref{sec:asymprncunc}, the asymptotic formula should give a good approximation when $t = O(N_c^{-1/2})$.  Indeed, as one can see from the graph, the asymptotic result is in agreement with the exact result for $t < 1/\sqrt{N_c} = 1/\sqrt{3} \approx 0.58$.

\newpage
\acknowledgments
We are grateful to Harold Widom and Estelle Basor for very useful correspondences.  Harold Widom also provided us with a note on analysis of Toeplitz determinant with a non-zero winding symbol to which we would like to express our thanks here.  

N.~M. would like to express his sincere gratitude to Amihay Hanany for teaching him Hilbert series as well as several useful tools/techniques and also for a long collaboration on various research projects.  He is grateful to the following institutes and collaborators for their very kind hospitality during the completion of this paper: Max-Planck-Institut f\"ur Physik (Werner-Heisenberg-Institut), Thomas Grimm, Sven Krippendorf (and his family), Oliver Schlotterer and Vudtiwat Ngampruetikorn.  He is very grateful to Aroonroj Mekareeya for his generosity in providing his laptop computer to use in this work. Finally, he would like to thank his family for the warm encouragement and support, as well as Thai taxpayers for funding his research via the DPST Project.

\appendix
\section{Multi-contour integrals and determinants} \label{app:haardet}
In this section, we show that one can rewrite the contour integrals in the Hilbert series of $U(N_c)$ SQCD in terms of a determinant.  A key tool is Gram's formula (see, \eg, Appendix A.12 of \cite{mehta}), which states as follows.  Let 
\bea
b_{ij} = \sum_{\alpha = 1}^n  v_{i \alpha} v^*_{j \alpha}~, \quad i, j =1, 2, \ldots,m~; \label{bdef}
\eea
then
\bea
\frac{1}{m!} \sum_{\alpha_1 =1}^n \cdots \sum_{\alpha_m =1}^n \left | \det( v_{i \alpha_j})_{1 \leq i, j \leq m} \right|^2  = \det (b_{ij})_{1 \leq i, j \leq m}~. \label{Gram}
\eea

\paragraph{The $U(N_c)$ Haar measure.} Let us first consider the $U(N_c)$ Haar measure
\bea
\int \ud \mu_{U(N_c)}  &=& \frac{1}{N_c! (2 \pi i)^{N_c}} \oint_{|z_1| =1}  \frac{\ud z_1}{z_1} \cdots \oint_{|z_{N_c}| =1}  \frac{\ud z_{N_c}}{z_{N_c}} | \Delta_{N_c} (z)|^2 \nn \\
&=&  \frac{1}{N_c! (2 \pi i)^{N_c}} \oint_{|z_1| =1}  \frac{\ud z_1}{z_1} \cdots \oint_{|z_{N_c}| =1}  \frac{\ud z_{N_c}}{z_{N_c}}\left | \det( z_a^{b-1})_{1 \leq a,b \leq N_c} \right |^2~. \label{haardet}
\eea
Let us now apply \eref{Gram}. We take $m = N_c$ and make the following substitutions:
\bea
\sum_{\alpha = 1}^n \quad &\longrightarrow& \quad \frac{1}{2 \pi i} \oint_{|z| =1} \frac{\ud z}{z} \nn \\
 \sum_{\alpha_1 =1}^n \cdots \sum_{\alpha_m =1}^n \quad &\longrightarrow& \quad \frac{1}{ (2 \pi i)^{N_c}} \oint_{|z_1| =1}  \frac{\ud z_1}{z_1} \cdots \oint_{|z_{N_c}| =1}  \frac{\ud z_{N_c}}{z_{N_c}}~, \nn \\
v_{i \alpha_j} \quad &\longrightarrow& \quad z_a^{b-1}~. \label{gramsub}
\eea
Thus, from \eref{bdef}, we find that
\bea
b_{ab} = \frac{1}{2 \pi i} \oint_{|z| =1} \frac{\ud z}{z} z^{a-1} (z^{b-1})^* = \frac{1}{2 \pi i} \oint_{|z| =1} \frac{\ud z}{z} z^{a-1} z^{1-b} =\frac{1}{2 \pi i} \oint_{|z| =1} \frac{\ud z}{z}  z^{a-b}~. \label{bdet}
\eea
Thus, from \eref{Gram}, \eref{haardet} and \eref{bdet}, we can rewrite the Haar measure of $U(N_c)$ as
\bea
\int \ud \mu_{U(N_c)}  = \det \left( \frac{1}{2 \pi i} \oint_{|z| =1} \frac{\ud z}{z}  z^{a-b} \right) _{1 \leq a,b \leq N_c}~. \label{haaruncdet}
\eea

\paragraph{The $U(N_c)$ SQCD Hilbert series.} Now consider the integral of the type:
\bea
\mathcal{I}(\xi) = \int \ud \mu_{U(N_c) (z_1, \ldots, z_{N_c})} \prod_{a=1}^{N_c} \phi(\xi, z_a)~. 
\eea
Note that this is also the integral we encounter in \eref{haarsym} in the context of $U(N_c)$ SQCD.  We make a similar substitution as in \eref{gramsub}:
\bea
\sum_{\alpha = 1}^n \quad &\longrightarrow& \quad \frac{1}{2 \pi i} \oint_{|z| =1} \frac{\ud z}{z} \phi(\xi,z)~, \nn \\
 \sum_{\alpha_1 =1}^n \cdots \sum_{\alpha_m =1}^n \quad &\longrightarrow& \quad \frac{1}{ (2 \pi i)^{N_c}} \oint_{|z_1| =1}  \frac{\ud z_1}{z_1} \cdots \oint_{|z_{N_c}| =1}  \frac{\ud z_{N_c}}{z_{N_c}} \prod_{a=1}^{N_c} \phi(\xi, z_a)~, \nn \\
v_{i \alpha_j} \quad &\longrightarrow& \quad z_a^{b-1}~. \label{gramsub2}
\eea
Thus, from \eref{Gram}, we find that
\bea
\mathcal{I}(\xi)  = \det \left( \frac{1}{2 \pi i} \oint_{|z| =1} \frac{\ud z}{z}  z^{a-b} \phi(\xi,z)  \right) _{1 \leq a,b \leq N_c}~.  \label{gramwithphi}
\eea

\section{A delta function in the contour integral} \label{sec:deltaintegral}
Let $f(z)$ a function which is analytic everywhere on the unit circle.  In this section, we show that the delta function in the integral
\bea
\oint_{|z|=1} \frac{\ud z}{z} \delta(z-1) f(z) 
\eea
can be replaced by an infinite sum as\footnote{We are very grateful to Harold Widom for pointing out this substitution to us.}
\bea \delta(z-1) \rightarrow \frac{1}{2 \pi i } \sum_{k=-\infty}^{\infty} z^k~,
\eea 
so that we have
\bea
\oint_{|z|=1} \frac{\ud z}{z} \delta(z-1) f(z) = \frac{1}{2 \pi i } \sum_{k=-\infty}^{\infty}  \oint_{|z|=1}  \frac{\ud z}{z} z^k f(z)~.
\eea

\paragraph{Proof.} Consider
\bea
f(1) =\oint_{|z|=1} \frac{\ud z}{z} \delta(z-1) f(z) = i \int_{-\pi}^\pi \ud \theta~\delta (e^{i \theta} -1) f(e^{i \theta}) ~.\label{diracdel}
\eea
Note that
\bea
\delta (e^{i \theta} -1) = -i \sum_{n= -\infty}^{\infty} \delta(\theta - 2\pi n) = -\frac{i}{2 \pi} \sum_{k=-\infty}^{\infty} e^{ik \theta}~, 
\eea
where the first equality can be verified by integrating from $2\pi m -\pi$ to $2\pi m+\pi$ (for any $m \in \BZ$) and noting that $-i = \int_{- \pi}^\pi \ud \theta~\delta (e^{i \theta} -1)$, whereas  in the last step we have used the Poisson summation formula.
Substituting this back into \eref{diracdel}, we obtain
\bea
\oint_{|z|=1} \frac{\ud z}{z} \delta(z-1) f(z) = \frac{1}{2 \pi} \sum_{k=-\infty}^{\infty} \int_{-\pi}^\pi \ud \theta~e^{i k \theta} f(e^{i \theta})  ~.
\eea
Writing $z = e^{i \theta}$ on the right hand side, we arrive at
\bea
\oint_{|z|=1} \frac{\ud z}{z} \delta(z-1) f(z) = \frac{1}{2 \pi i } \sum_{k=-\infty}^{\infty}  \oint_{|z|=1}  \frac{\ud z}{z} z^k f(z)~.
\eea
This amounts to the replacement 
\bea \delta(z-1) \rightarrow \frac{1}{2 \pi i } \sum_{k=-\infty}^{\infty} z^k~. 
\eea

\end{document}